\documentclass[a4paper,aps,prd,twocolumn,nofootinbib]{revtex4}
\usepackage[latin9]{inputenc}
\usepackage{hyperref}
\hypersetup{
  colorlinks=true,        
  linkcolor=blue,         
  citecolor=cyan,         
}
\usepackage{breakurl}
\usepackage{graphicx}
\usepackage{epsf}
\usepackage{epsfig}
\usepackage{amssymb,amsmath}
\usepackage[usenames]{color}
\usepackage{amssymb}
\usepackage{times}
\usepackage{comment}

\newcommand\snn{\sqrt{s_\text{NN}}}

\newcommand\pt{p_\text{T}}

\newcommand\raa{R_\text{AA}}

\newcommand\dbar{\bar{D}^{0}}
\newcommand\dv {\Delta v_{1}}



\begin{document}

\title{Probing the initial longitudinal density profile and  electromagnetic field in ultrarelativistic heavy-ion collisions with heavy quarks}
\author{Ze-Fang Jiang$^{~1,2}$}
\email{jiangzf@mails.ccnu.edu.cn}
\author{Shanshan Cao$^{~3}$}
\email{shanshan.cao@sdu.edu.cn}
\author{Wen-Jing Xing$^{~2}$}
\author{Xiang-Yu Wu$^{~2}$}
\author{C. B. Yang$^{~2}$}
\author{Ben-Wei Zhang$^{~2,4}$}
\affiliation{$^1$ Department of Physics and Electronic-Information Engineering, Hubei Engineering University, Xiaogan, Hubei, 432000, China}
\affiliation{$^2$ Institute of Particle Physics and Key Laboratory of Quark and Lepton Physics (MOE), Central China Normal University, Wuhan, Hubei, 430079, China}
\affiliation{$^3$ Institute of Frontier and Interdisciplinary Science, Shandong University, Qingdao, Shandong, 266237, China}
\affiliation{$^4$ Guangdong Provincial Key Laboratory of Nuclear Science, Institute of Quantum Matter,
South China Normal University, Guangzhou, Guangdong, 510006, China}

\begin{abstract}
Heavy quarks are valuable probes of the electromagnetic field and the initial condition of the quark-gluon plasma (QGP) matter produced in high-energy nuclear collisions. Within an improved Langevin model that is coupled to a (3+1)-dimensional viscous hydrodynamic model, we explore the origin of the directed flow coefficient ($v_1$) of heavy mesons and their decay leptons, and its splitting ($\dv$) between opposite charges. We find that while the rapidity dependence of the heavy flavor $v_1$ is mainly driven by the titled energy density profile of the QGP with respect to the longitudinal direction at the RHIC energy, it is dominated by the electromagnetic field at the LHC energy. The $\dv$ serves as a novel probe of the spacetime evolution profile of the electromagnetic field. Our results of $D$ mesons and their decay electrons are consistent with the available data at RHIC and LHC, and our predictions on the heavy flavor decay muons can be further tested by future measurements.

\end{abstract}
\maketitle
\date{\today}

\section{Introduction}
\label{emsection1}

Heavy-ion collisions provide a unique opportunity to study the color deconfined state of nuclear matter, known as the Quark-Gluon Plasma (QGP)~\cite{Shuryak:2014zxa}. Heavy flavor spectra are among the cleanest observables that reveal the QGP properties probed at different energy scales~\cite{Dong:2019byy,Dong:2019unq}. Due to their large masses, heavy quarks are mostly produced in the very early stage of high-energy nuclear collisions, and then interact with the nuclear medium with their flavors conserved before hadronizing into heavy flavor hadrons on the QGP boundary, thus performing a tomography of the entire evolution history of the QGP.

Tremendous efforts have been devoted in both experimental~\cite{STAR:2014wif,ALICE:2015vxz,PHENIX:2006iih,ALICE:2017pbx} and theoretical~\cite{Moore:2004tg,He:2012df,Uphoff:2012gb,Nahrgang:2014vza,Song:2015sfa,Cao:2015hia,Cao:2016gvr,Prado:2016szr,Prino:2016cni,Liu:2016zle,Zhou:2016wbo,Cao:2017hhk,Scardina:2017ipo,Beraudo:2017gxw,Ke:2018tsh,Cao:2018ews,Xing:2019xae,Li:2020kax,Li:2021xbd} studies on the nuclear modification of heavy quarks inside the QGP, including the suppression factor $\raa$ and the elliptic flow coefficient $v_{2}$ of heavy flavor hadrons and their decay leptons. This allows one to acquaint with the mass and flavor dependence of jet-medium interaction at high transverse momentum ($p_\mathrm{T}$) and the thermalization process of heavy quarks at low $p_\mathrm{T}$. Another important aspect of heavy quark study is utilizing them to probe the hadronization process from the quark-gluon state to the hadronic state of nuclear matter. This can be reflected by the heavy flavor hadron chemistry, such as the enhancement of $\Lambda_c/D^0$, $D_s/D^0$ and $B_s/B^+$ ratios in nucleus-nucleus collisions with respect to proton-proton collisions. Relative investigations have recently been improved from both experimental~\cite{ALICE:2018lyv,ALICE:2018hbc,CMS:2018eso,CMS:2019uws,STAR:2019ank,STAR:2021tte} and theoretical~\cite{Plumari:2017ntm,He:2019vgs,Cho:2019lxb,Cao:2019iqs} sides.

While higher-order harmonic (elliptic $v_2$, triangular $v_3$, etc.) flow coefficients mainly characterize the heavy quark energy loss and thermalization through an asymmetric medium in the transverse plane, the rapidity dependence of directed flow ($v_1$) focuses more on the asymmetry in the reaction plane of heavy-ion collisions and becomes a novel tool to probe the longitudinal distribution of the initial profile of the QGP. It has been proposed that the heavy flavor hadron $v_1$ could be more than an order of magnitude larger than that of the light hadrons emitted from the QGP~\cite{Chatterjee:2017ahy,Chatterjee:2018lsx,Nasim:2018hyw}, which has soon been confirmed by the STAR data~\cite{STAR:2019clv}. More detailed studies have later been performed in Refs.~\cite{Oliva:2020doe,Beraudo:2021ont} using transport models of heavy quarks coupled to a QGP medium that takes into account the initial longitudinal tilt~\cite{Bozek:2010bi} in the reaction plane with respect to the beam axis.

Another crucial origin of the heavy flavor $v_1$ is the strong electromagnetic fields produced in heavy-ion collisions. It has been estimated that the magnetic field in the early stage of nuclear collisions ($<0.5$~fm/$c$) can reach several times of $10^{18}$~Gauss in Au+Au collisions at RHIC and $10^{19}$~Gauss in Pb+Pb collisions at LHC~\cite{Fukushima:2008xe,Bzdak:2011yy,Deng:2012pc,Zhong:2014cda,Li:2016tel,Zhao:2017nfq}. Such strong electromagnetic field can deflect the motion of heavy quarks traversing the medium, causing the separation of $v_1$ between $D^{0}$ and $\dbar$ mesons in the end. Interestingly, while the STAR measurement~\cite{STAR:2019clv} observes decreasing $v_1$ with respect to rapidity ($y$) for both $D^{0}$ and $\dbar$, with very small difference between their magnitudes, the ALICE measurement~\cite{ALICE:2019sgg} presents apparent splitting of the directed flow ($\dv$) between opposite charges, with $D^{0}$ increasing and $\dbar$ decreasing with pseudorapidity ($\eta$). This puzzling observation has attracted a series of investigations on heavy quark dynamics in the presence of electromagnetic field~\cite{Das:2016cwd,Chatterjee:2018lsx,Oliva:2020doe,Sun:2020wkg}.

The different behaviors of the heavy meson $v_1$ observed at STAR and ALICE suggest different competing effects between the asymmetric medium and the electromagnetic field at RHIC and LHC. Based on the pioneer studies above, we conduct a systematic exploration of the origin of the heavy flavor $v_1$ at different colliding energies in this work. The heavy quark evolution through the QGP is described using a modified Langevin approach~\cite{Cao:2013ita,Cao:2015hia} that incorporates the thermal diffusion of heavy quarks inside the QGP, medium-induced gluon emission, as well as the Lorentz force due to the electromagnetic field. With the tilted geometry of the initial energy density distribution with respect to the longitudinal direction~\cite{Jiang:2021foj,Jiang:2021ajc}, the spacetime evolution profile of the QGP is simulated with the (3+1)-D viscous hydrodynamic model CLVisc~\cite{Pang:2012he,Pang:2018zzo,Wu:2018cpc,Wu:2021fjf}. Within this sophisticated framework, we find that the heavy meson $v_1$ is dominated by the heavy quark interaction with the tilted QGP medium at the RHIC energy, while by the heavy quark interaction with the electromagnetic field at the LHC energy. By comparing between two different model calculations of the electromagnetic field, we also find that the $\dv$ between $D^0$ and $\dbar$ is sensitive to the evolution profile of the field. These findings are further confirmed with our predictions on the heavy flavor decay electrons and muons.

This work will be organized as follows. In Sec.~\ref{section2}, we will briefly review our modelings of the tilted initial condition of the bulk medium and its subsequent evolution via the CLVisc hydrodynamic simulation, and two different setups of the electromagnetic field. In Sec.~\ref{section3}, we will develop our modified Langevin approach that describes the heavy quark interaction with both the QGP medium and the external field. Our numerical results on the heavy flavor $v_1$ and $\dv$ will be presented and discussed in Sec.~\ref{section4}. In the end, we summarize in Sec.~\ref{section5}.

\section{Spacetime evolution of the QGP and the electromagnetic field}
\label{section2}
\subsection{Hydrodynamic evolution of the QGP}
\label{subsec:hydro}

Before studying the heavy quark interaction with the QGP medium, we first discuss the evolution of the QGP fireballs within the (3+1)-D viscous hydrodynamic model CLVisc~\cite{Pang:2018zzo} coupled with the tilted initial energy density distribution in the reaction plane of non-central heavy-ion collisions~\cite{Jiang:2021foj,Jiang:2021ajc}.

The initial energy density $\varepsilon(x,y,\eta_{s})$ is given by~\cite{Pang:2018zzo}
\begin{equation}
\begin{aligned}
\varepsilon(x,y,\eta_{s})=K \cdot W(x,y,\eta_{s}) \cdot H(\eta_{s}),
\label{eq:ekw}
\end{aligned}
\end{equation}
where $K$ is an overall normalization factor determined by the soft particle yield in different collision systems, $(x,y)$ represents the transverse plane, and $\eta_s$ is the spacetime rapidity. The total weight function $W(x,y,\eta_{s})$ is defined as
\begin{equation}
\begin{aligned}
W(x,y,\eta_{s})=\frac{(1-\alpha)W_\text{N}(x,y,\eta_{s})+\alpha n_\text{BC}(x,y)}{\left[(1-\alpha)W_\text{N}(0,0,0)+\alpha n_\text{BC}(0,0)\right]|_{\mathbf{b}=0}},
\label{eq:wneta}
\end{aligned}
\end{equation}
where $W_\mathrm{N}$ represents the weight contributed by wounded nucleons, $n_\text{BC}$ represents contributions from binary collisions~\cite{Jiang:2021ajc}, and the collision hardness parameter $\alpha$ measures the relative contributions between them. Following our recent studies~\cite{Jiang:2021foj,Jiang:2021ajc}, the asymmetric distribution with respect to the beam axis is introduced into $W_\text{N}$ as
\begin{equation}
\begin{aligned}
W_\text{N}(x,y,\eta_{s})=&[T_{1}(x,y)+T_{2}(x,y)]\\
+&H_{t}[T_{1}(x,y)-T_{2}(x,y)]\tan\left(\frac{\eta_{s}}{\eta_{t}}\right),
\label{eq:mnccnu}
\end{aligned}
\end{equation}
in which $T_1$ and $T_2$, containing the nucleon-nucleon inelastic scattering cross section $\sigma_\text{NN}$, are the density distribution of participant nucleons from the two colliding nuclei traveling in the positive and negative $z$ directions respectively~\cite{Jiang:2021ajc}; and $H_t\tan (\eta_s/\eta_t)$ reflects the strength of imbalance between particle emission in the forward and backward spacetime rapidities ($\eta_s$) along the direction of impact parameter ($\textbf{b}$). In the present work, we adopt $H_{t}=3.9$ for 10-80\% Au+Au collisions at $\snn=200$~GeV and $H_{t}=0.70$ for 10-40\% Pb+Pb collisions at $\snn=5.02$~TeV; $\eta_{t}=8.0$ is used for both systems. These values have been adjusted in Ref.~\cite{Jiang:2021ajc} for a satisfactory description of the soft hadron $v_1$ in their corresponding collision systems. Additionally, in Eq.~(\ref{eq:ekw}), a function
\begin{equation}
\begin{aligned}
H(\eta_{s})=\exp\left[-\frac{(|\eta_{s}|-\eta_{w})^{2}}{2\sigma^{2}_{\eta}}\theta(|\eta_{s}|-\eta_{w}) \right]
\label{eq:heta}
\end{aligned}
\end{equation}
is introduced to describe the plateau structure of the rapidity distribution of emitted hadrons at mid-rapidity, in which $\eta_{w}$ determines the width of the central rapidity plateau while $\sigma_{\eta}$ determines the width (speed) of the Gaussian decay away from the plateau region~\cite{Pang:2018zzo}. Related model parameters are summarized in Tab.~\ref{t:modelparameters}.

\begin{figure}[tbp]
\begin{center}
\includegraphics[trim=0cm 0.0cm 0cm 0cm,width=8 cm,height=6 cm,clip]{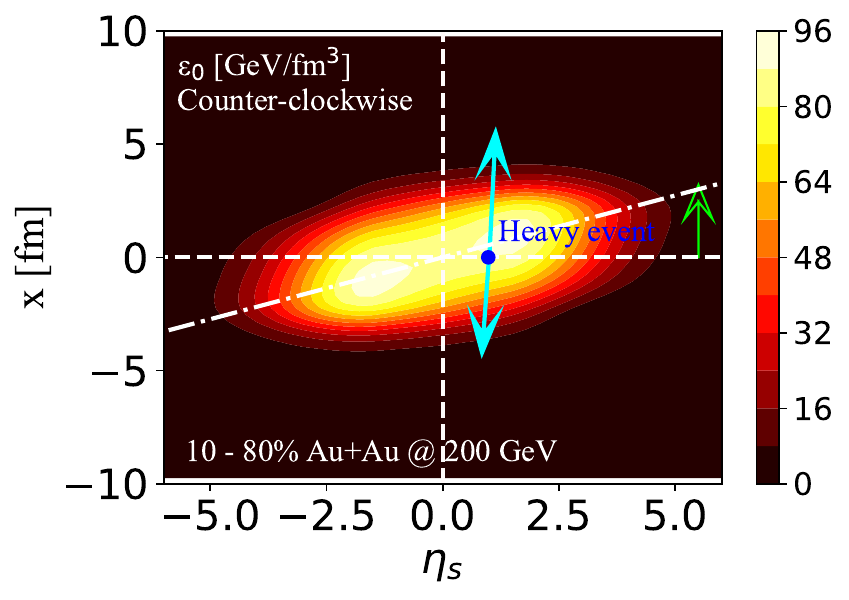}~~~\\
\end{center}
\caption{(Color online) The initial energy density of the bulk medium in the $\eta_{s}$-$x$ plane at $\tau_0 = 0.2$~fm/$c$ in 10-80\% ($b = 8.5$~fm) Au+Au collisions at $\sqrt{s_\text{NN}}=200$~GeV. The empty arrow (limes color) illustrates the counter-clockwise tilted geometry with respect to the longitudinal direction, while the solid arrows (aqua color) sketch the heavy quark propagation through the medium.}
\label{fig:tiltedMedium}
\end{figure}

\begin{table}[!h]
\begin{center}
\begin{tabular}{ |c| c |c|  }
\hline
~~~        & Au+Au            & Pb+Pb          \\
           & $\snn$ = 200 GeV & $\snn$ = 5.02 TeV  \\
\hline
$\tau_{0}$ (fm/$c$)             & 0.2        & 0.2              \\
\hline
 $K$ (GeV/fm$^{3}$)                  & 125.0       & 490.0            \\
\hline
$\alpha$                      & 0.05       & 0.05       \\
\hline
$\sigma_\text{NN}$ (mb)                 & 42      & 68       \\
\hline
$\eta_{w}$                    & 1.3      & 2.2       \\
\hline
$\sigma_{\eta}$               & 1.5      & 1.8     \\
\hline
\end{tabular}
\caption{\label{t:modelparameters} Model parameters for the initial energy density distributions at RHIC and LHC~\cite{Pang:2018zzo,Loizides:2017ack}.}
\end{center}
\end{table}

The initial fluid velocity at $\tau_{0}$ is assumed to follow the Bjorken approximation in this work as $v_{x} = v_{y} =0$ and $v_{z} = z/t$, where the initial transverse expansion and the asymmetric distribution of $v_z$ along the impact parameter ($x$) direction are ignored. More sophisticated initial velocity profiles will be studied in an upcoming effort.

With these setups, the initial energy density distribution is illustrated in Fig.~\ref{fig:tiltedMedium}, where a counter-clockwise tilted geometry in the $x$-$\eta_s$ plane with respect to the longitudinal direction can be seen. This tilted initial condition was shown essential for understanding the directed flow of soft hadrons emitted from the QGP~\cite{Jiang:2021ajc}. Since heavy quarks are produced from the initial hard scatterings within the overlapping region between the two colliding nuclei, their initial spatial distribution is expected to be symmetric about the $y$-$\eta_s$ plane. The tilted medium above then give rise to a longer path length (stronger energy loss) of heavy quarks traveling along $+x$ than $-x$ direction towards the $+z$ region, leading to a negative $x$-component of the average heavy quark momentum ($\langle p_x\rangle$), thus a negative $v_1$. The reverse is expected for heavy quarks propagating towards the $-z$ region.

With this initial condition, the subsequent evolution of the bulk medium follows the hydrodynamic equations as~\cite{Jiang:2020big,Jiang:2018qxd,Denicol:2012cn,Romatschke:2009im,Romatschke:2017ejr}
\begin{equation}
\partial_{\mu}T^{\mu\nu}=0,
\label{eq:tmn}
\end{equation}
where the energy-momentum tensor is given by
\begin{equation}
T^{\mu\nu}=\varepsilon u^{\mu}u^{\nu}-(P+\Pi)\Delta^{\mu\nu} + \pi^{\mu\nu},
\label{eq:tensor}
\end{equation}
which involves the local energy density $\varepsilon$,
the fluid four-velocity $u^{\mu}$, the pressure $P$, the bulk viscosity pressure $\Pi$ and the shear viscosity tensor $\pi^{\mu\nu}$.
The projection tensor is defined as $\Delta^{\mu\nu} = g^{\mu\nu}-u^{\mu}u^{\nu}$ with the metric tensor $g^{\mu\nu} = \text{diag} (1,-1,-1,-1)$.
The hydrodynamic equations are solved together with the lattice QCD equation of state (EoS) from the Wuppertal-Budapest group~\cite{Borsanyi:2013bia}.
For a minimal model, a constant shear-viscosity-to-entropy-density ratio is taken as $\eta_{v}/s = 0.08$ ($\eta_{v}$ for the shear viscosity), while the bulk viscosity and the net baryon density are ignored in the current work.
With these setups, our hydrodynamic calculation is able to provide a satisfactory description of the soft hadron spectra, including their pseudorapidity-dependent yield ($dN_{\textrm{ch}}/d\eta$) and directed flow coefficient ($v_{1}$)~\cite{Pang:2018zzo,Jiang:2021ajc,Jiang:2021foj}.

\subsection{Time evolution of electromagnetic field}
\label{subsec:emfield}

Intensive studies have been performed in the past decade on the strong electromagnetic field generated in relativistic heavy-ion collisions.
Although the evaluation of the field at the initial time of nuclear collisions ($t=0$) has been settled in earlier work~\cite{Deng:2012pc,McLerran:2013hla}, how it evolves with spacetime is still an open question~\cite{McLerran:2013hla,Gursoy:2014aka,Tuchin:2015oka,Inghirami:2019mkc}.
The challenges come from the complicated medium environment that starts from an extremely non-equilibriated condition and rapidly evolves from the Color Glass Condensate (CGC) state to the QGP state.
While lattice QCD calculations can provide the electric conductivity $\sigma_{\textrm{el}}$ of the QGP medium, large uncertainties still remain~\cite{Ding:2010ga}.
In this work, we will employ two different model calculations of the spacetime profiles of the electromagnetic field and compare their impacts on the heavy flavor $v_1$. Following earlier studies~\cite{Gursoy:2014aka,Das:2016cwd,Chatterjee:2018lsx,Sun:2020wkg,Oliva:2020doe}, only the two dominant components, $E_{x}$ and $B_{y}$, are included in our calculation.

\begin{figure}[tbp]
\begin{center}
\includegraphics[trim=0cm 0.0cm 0cm 0cm,width=8 cm,height=6 cm,clip]{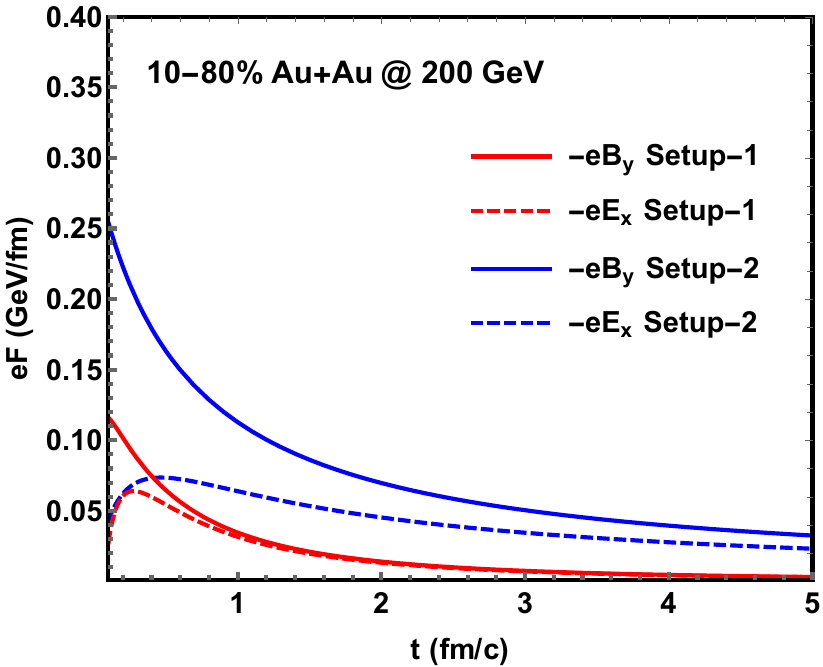}~~~\\
\includegraphics[trim=0cm 0.0cm 0cm 0cm,width=7.8 cm,height=6 cm,clip]{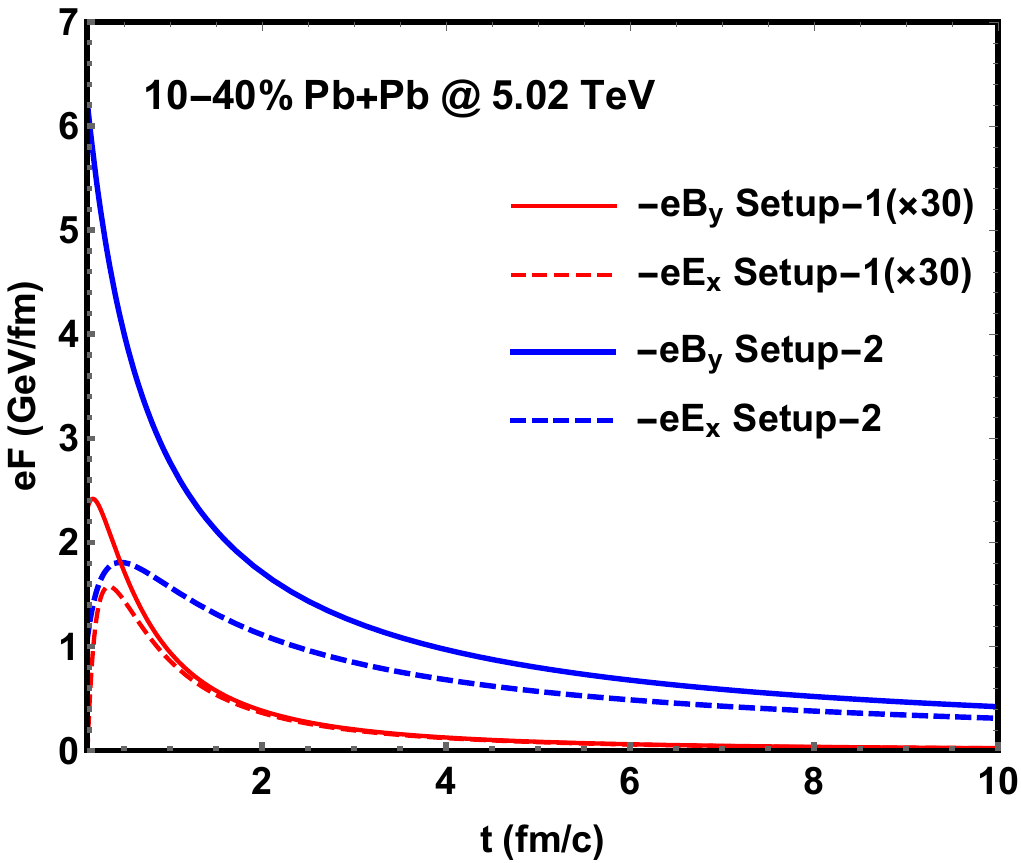}~
\end{center}
\caption{(Color online) Time evolution of the electromagnetic field at $x=y=0$ and $\eta_{s}=1.0$, compared between two different model setups, upper panel for 10-80\% Au+Au collisions at $\snn=200$~GeV, and lower panel for 10-40\% Pb+Pb collisions at $\snn= 5.02$~TeV.}
\label{f1:EBfield}
\end{figure}

\textbf{Setup-1} The first setup of electromagnetic field is based on the solution of the Maxwell's equations with the source of moving charges contributed by the spectators in nuclear collisions~\cite{Tuchin:2015oka,Das:2016cwd,Gursoy:2014aka}. In most calculations, an instantaneous thermalization within the overlapping region of collisions is assumed, and a constant electric conductivity  $\sigma_{\textrm{el}}=0.023$~fm$^{-1}$ is adopted from the lattice QCD evaluations~\cite{Ding:2010ga,Amato:2013naa}. Although introducing conductivity slows down the decay of electromagnetic field, it significantly reduces the strength of the field at early time (before the realistic starting time of the QGP) compared to the vacuum environment~\cite{Deng:2012pc,McLerran:2013hla,Li:2016tel}.

\textbf{Setup-2} The second setup is adopted from Ref.~\cite{Sun:2020wkg}, where the magnetic field at the medium center is initialized with the value calculated in vacuum: $B_{y}(t=x=y=z=0) = - B_{0}$, with $eB_{0}$ taken as 0.06~GeV$^{2}$ ($\approx 2.97 m_{\pi}^{2}$) for RHIC and 1.43~GeV$^{2}$ ($\approx 73 m_{\pi}^{2}$) for LHC ~\cite{Deng:2012pc,Yin:2015fca,Jiang:2016wve,Pang:2016yuh}. Its spacetime distribution is then modeled as
\begin{equation}
\begin{aligned}
eB_{y}(\tau,x,y)&=-eB_{0}\rho(\tau)\rho_{B}(x,y),
\label{eq:eB}
\end{aligned}
\end{equation}
in which $\rho(\tau) = 1/(1+\tau/\tau_{B})$ with $\tau_{B}=0.4$~fm/$c$ provides the evolution with respect to the proper time ($\tau$)~\cite{Sun:2020wkg}, and $\rho_{B}(x,y)=\exp\left[-x^{2}/(2\sigma_{x}^{2})-y^{2}/(2\sigma_{y}^{2})\right]$ provides the spatial distribution with  $\sigma_{x}$ and $\sigma_{y}$ being the Gaussian widths along the $x$ and $y$ directions~\cite{Roy:2017yvg}. Boost invariance is assumed for the field strength at different spacetime rapidites ($\eta_s$).

With the magnetic field given above, the $eE_{x}$ can be determined by solving the Faraday's Law $\nabla\times \textrm{\textbf{E}} = -\partial \textrm{\textbf{B}}/\partial t$ as
\begin{equation}
\begin{aligned}
eE_{x}(t,x,y,\eta_{s}) &= e B_0 \rho_{B}(x,y)\\
&\times\int_{0}^{\eta_{s}}d\chi \rho'\left(\frac{t}{\cosh\chi}\right)\frac{t}{\cosh\chi},
\label{eq:eE}
\end{aligned}
\end{equation}
where $\rho'$ denotes the derivative of $\rho(\tau)$ with respect to $\tau$.

This modeling of the electromagnetic field is expected to be applicable when $\eta_s$ and the transverse coordinate ($\sqrt{x^{2}+y^{2}}$) are not large. Otherwise, one needs to solve the full Maxwell equations with complex boundary conditions~\cite{Sun:2021joa}.

Shown in Fig.~\ref{f1:EBfield} are the time evolution of $eE_{x}$ and $eB_{y}$, compared between our two model setups, in 10-80\% (impact parameter $b=8.54$~fm) Au+Au collisions at $\snn=200$~GeV and 10-40\% ($b=7.65$~fm) Pb+Pb collisions at $\snn=5.02$~TeV. Results are shown for the position at $(x,y) = (0,0)$
and $\eta_{s}$ = 1.0. One observes that for $t> 1 $~fm/$c$, the magnitude of $eB_{y}$ is larger than $eE_{x}$ in setup-2, but almost indistinguishable in setup-1. This will affect the direction of the deflection of charged particles and in the end the sign of the heavy flavor $v_1$. In addition, the maximum value of $eB_{y}$ in setup-1 is approximately 30 times smaller than that in setup-2 (vacuum value) in 5.02~ATeV Pb+Pb collisions, while 2 times smaller in 200~AGeV Au+Au collisions. For numerical simulations of heavy quarks, the grid range of the electromagnetic field will be assigned as 0.2~fm/$c< \tau < 10.2$~fm/$c$, $0 < x_\mathrm{T} <10.2$~fm, $0 < \phi < 2\pi$ and $-4.0 < \eta_{s} < 4.0$ in this work, with $x_\mathrm{T}$ and $\phi$ being the radius and azimuthal angle in the transverse plane.

\section{Transport of heavy quarks}
\label{section3}

In this work, we further develop the modified Langevin approach~\cite{Cao:2013ita,Cao:2015hia} to simultaneously describe the heavy quark interaction with the QGP and the electromagnetic field. The modified Langevin equation is now expressed as
\begin{equation}
\begin{aligned}
\frac{d\vec{p}}{dt}=-\eta_\mathrm{D}(\vec{p})\vec{p}+\vec{\xi}+\vec{f}_{g}+q(\vec{E}+\vec{v}\times\vec{B}).
\label{eq:CaoLangevin}
\end{aligned}
\end{equation}
The first two terms on the right hand side represent the drag force and thermal random force on heavy quarks inside a thermal medium. The third term $\vec{f}_{g}$ provides the recoil force experienced by heavy quarks when they emit medium-induced gluons. And the last term is introduced for the Lorentz force on heavy quarks in the presence of electromagnetic field.

For quasielastic scatterings, the minimal assumption of the momentum ($\vec{p}$) independent $\vec{\xi}$ is adopted. It is determined by the white noise $\big\langle \xi^i (t) \xi^j(t') \big\rangle = \kappa \delta^{ij} \delta(t-t')$ where $\kappa$ is known as the the momentum space diffusion coefficient which is related to the drag coefficient via the fluctuation-dissipation relation $\eta_\mathrm{D}(p)=\kappa/(2TE)$ with $T$ and $E$ being the medium temperature and heavy quark energy respectively. The spatial diffusion coefficient of heavy quarks is then given by $D_\mathrm{s}\equiv T/[M\eta_\mathrm{D}(0)]=2T^2/\kappa$, in which $M$ denotes the mass of heavy quarks.

The recoil force is given by $f_g=d\vec{p}_g/dt$, where $\vec{p}_g$ represents the momentum of medium-induced gluons, whose spectrum is taken from the higher-twist energy loss formalism~\cite{Guo:2000nz,Majumder:2009ge,Zhang:2003wk}. The strength of this term is controlled by the jet quenching parameter $\hat{q}$, which is related to the momentum space diffusion coefficient via the dimension factor $\hat{q}=2\kappa$. For detailed implementation, one may refer to our previous work Ref.~\cite{Cao:2015hia}. And the systematic uncertainties from various model ingredients have been discussed in Ref.~\cite{Li:2020kax}. By convention, we take $D_\mathrm{s}$ as the input parameter for our model calculation, whose value is set as $D(2\pi T) = 4.0$ at RHIC and $D(2\pi T) = 7.0$ at LHC for a reasonable description of the observed nuclear modification factors of heavy mesons. Note that the heavy quark transport coefficient quoted in the present study only measures the strength of the thermal random force from the QGP medium, while effects from the electromagnetic field is treated as an external force in Eq.~(\ref{eq:CaoLangevin}). In principle, the total transport coefficient should also include contribution from the electromagnetic field. 

The spatial distributions of heavy quarks are initialized using the binary collision positions from the Monte-Carlo Glauber model, while their momentum spectra are calculated using the leading-order perturbative QCD calculation that includes pair production and flavor excitation processes, coupled to the CTEQ6 parton distribution function~\cite{Kretzer:2003it} and EPS09 parametrization of nuclear shadowing effect~\cite{Eskola:2009uj} in nucleus-nucleus collisions. In the present study, we assume heavy quarks start interacting with the medium since the initial time of the hydrodynamic evolution ($\tau_0=0.2$~fm/$c$).
Upon traveling across the QGP boundary with a decoupling temperature set as $T_\text{d} = 165$~MeV, heavy quarks are converted to heavy flavor mesons via our hybrid fragmentation and coalescence model~\cite{Cao:2019iqs} that is well constrained by the heavy flavor hadron chemistry measured at RHIC and LHC. In the end, these heavy flavor hadrons decay into leptons via Pythia simulation~\cite{Sjostrand:2006za}.

\section{Nuclear modification and directed flow of heavy mesons and their decay leptons}
\label{section4}

In this section, we present our numerical results on the heavy flavor observables and discuss how they are affected by the initial geometry of the QGP and the evolution profiles of the electromagnetic field. We will concentrate on two main observables, nuclear modification factor $R_\text{AA}$ and the directed flow coefficient $v_1$. The former is defined as the ratio of the particle spectra between nucleus-nucleus collisions and proton-proton collisions,
\begin{equation}
\begin{aligned}
R_{\textrm{AA}}=\frac{1}{\mathcal{N}}\frac{dN_{\textrm{AA}}/dydp_{\textrm{T}}}{dN_{\textrm{pp}}/dydp_{\textrm{T}}},
\label{eq:raa}
\end{aligned}
\end{equation}
where $\mathcal{N}$ is the average number of binary nucleon-nucleon collisions in a given setup of nucleus-nucleus collisions. The directed flow is the first-order Fourier coefficient of the angular distribution of the particle spectra and can be obtained via
\begin{equation}
\begin{aligned}
v_{1}=\left\langle \cos(\phi-\Psi_1)\right\rangle=\left\langle\frac{p_{x}}{\pt}\right\rangle,
\label{eq:v1}
\end{aligned}
\end{equation}
where $\Psi_1$ is the first-order event plane angle and $\langle...\rangle$ denotes the average over the final-state heavy mesons or their decay leptons obtained from our Langevin simulation. Since we use the optical Glauber model, as described in Sec.~\ref{subsec:hydro}, to initialize the energy density distribution of the QGP, event-by-event fluctuations have not been taken into account in this work. Therefore, the event plane in the final state here is the same as the participant plane in the initial state, which is also the same as the spectator plane determined using the deflected neutrons in realistic experimental measurements. More sophisticated analysis needs to be implemented in our future work after introducing the event-by-event fluctuation.

\subsection{$\raa$ and $v_{1}$ of heavy mesons}
\label{section4-1}

We start with the nuclear modification factor and directed flow coefficient of $D$ mesons, including the splitting of $v_1$ between $D^{0}$ and $\dbar$, in 200~AGeV Au+Au collisions at RHIC and 5.02~AGeV Pb+Pb collisions at LHC.

\begin{figure}[tbp]
\begin{center}
\includegraphics[trim=0cm 0.2cm 0cm 0cm,width=8 cm,height=5 cm,clip]{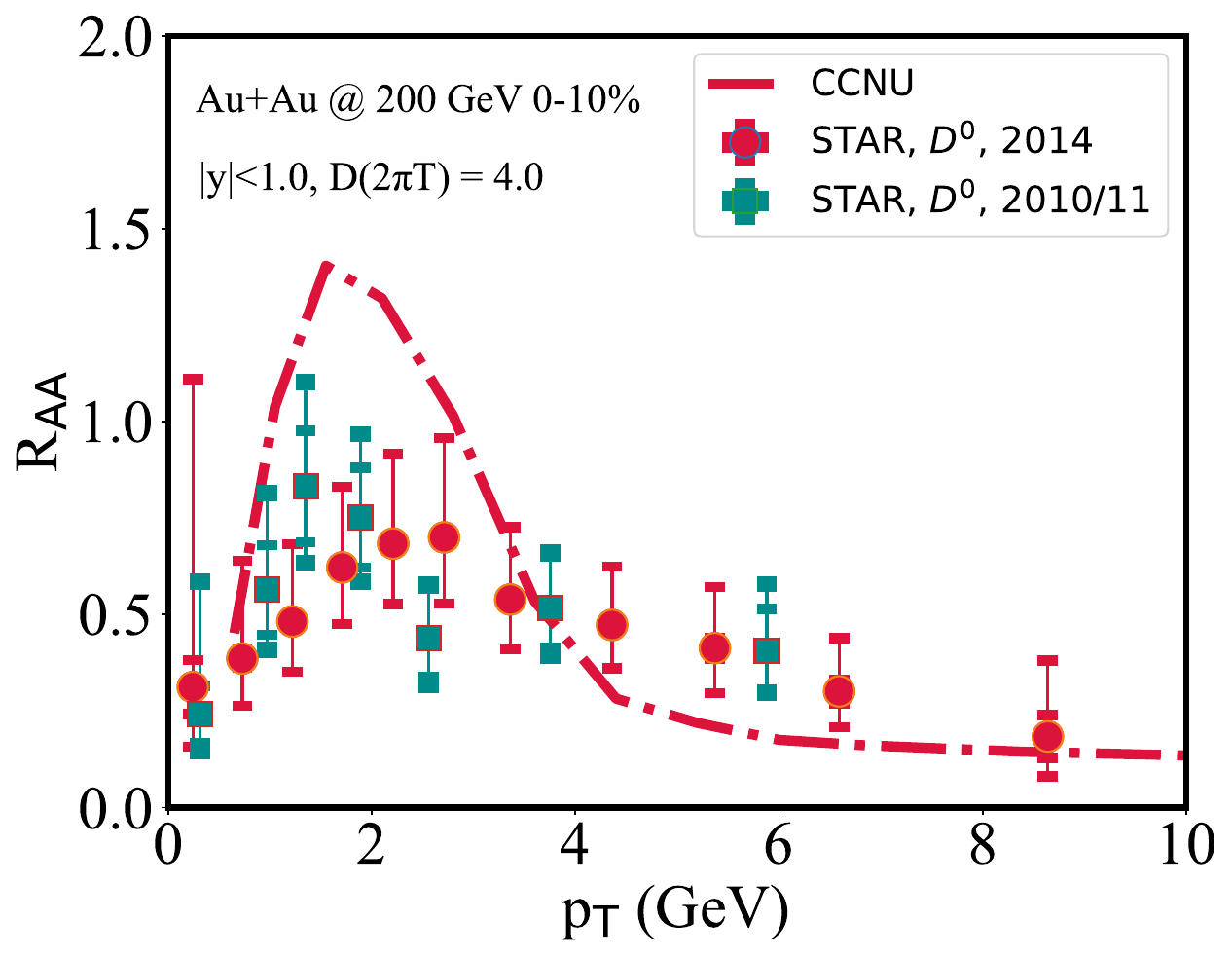}~\\
\includegraphics[trim=0cm 0.2cm 0cm 0cm,width=8 cm,height=5 cm,clip]{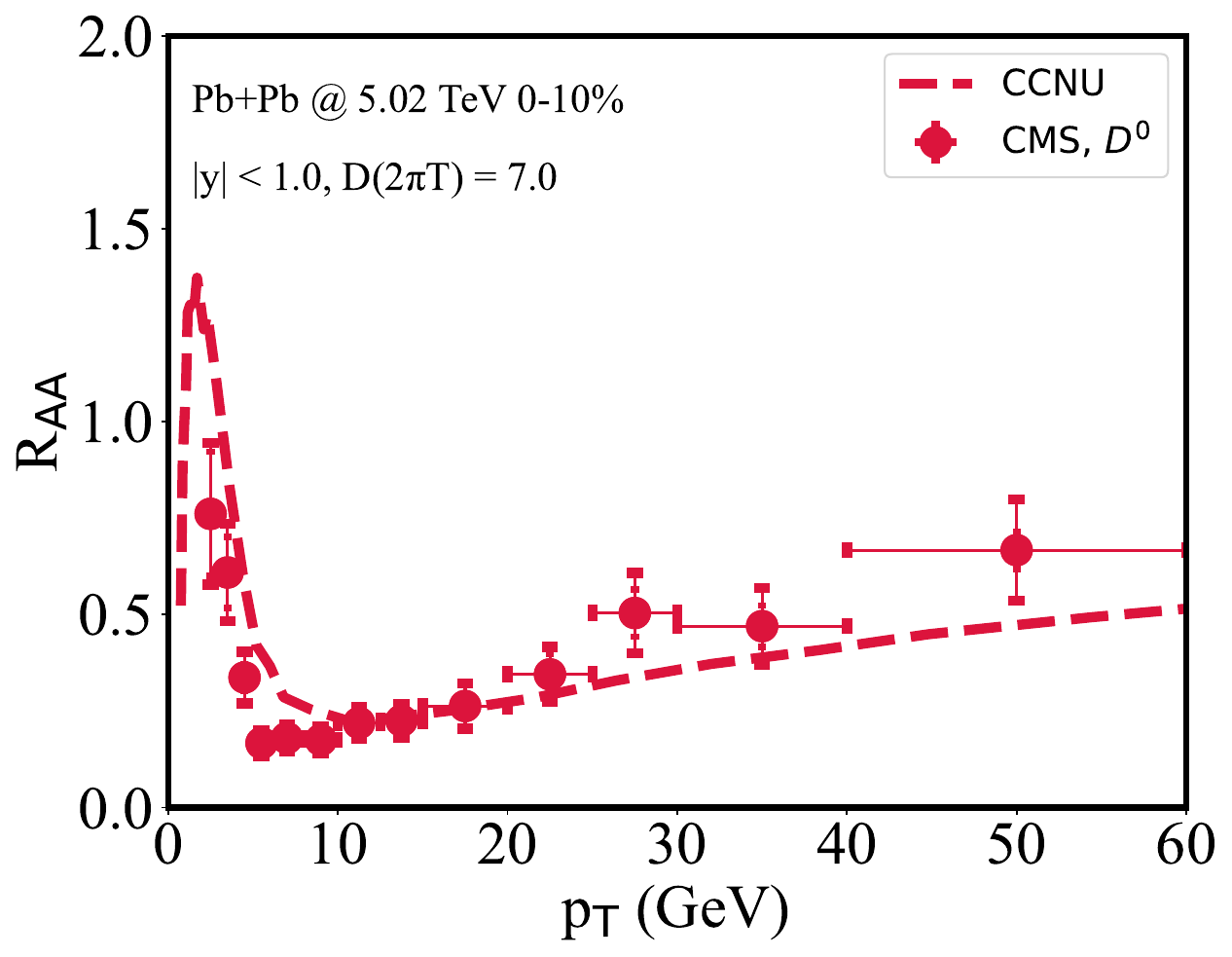}~
\end{center}
\caption{(Color online) Nuclear modification factor of $D$ mesons in 0-10\% 200~AGeV Au+Au collisions (upper panel) and 5.02~ATeV Pb+Pb collisions (lower panel), compared to the STAR~\cite{Radhakrishnan:2019gbl} and CMS~\cite{CMS:2017qjw} data respectively.}
\label{F1:RAA_AuAu}
\end{figure}

In Fig.~\ref{F1:RAA_AuAu}, we first present the $\raa$ of $D$ mesons in central Au+Au and Pb+Pb collisions. Results are shown for the mid-rapidity region ($|y|<1$) and compared to experimental data at RHIC and LHC. With the spatial diffusion coefficient set as $D(2\pi T)=4$ at RHIC and 7 at LHC, a reasonable agreement has been obtained between our calculation and the data. A larger value of the diffusion can be understood with a weaker average interaction strength between heavy quarks and the hotter QGP matter at LHC than at RHIC. The peak structure of the $D$ meson $\raa$ results from the coalescence process when charm quarks hadronize~\cite{Cao:2015hia}. Model uncertainties still remain in this non-perturbative process. We have verified that effect of the electromagnetic field is negligible in the heavy meson $R_\mathrm{AA}$, considering the much weaker Lorentz force compared to the elastic and inelastic scatterings between heavy quarks and the QGP medium. A reasonable description of the $D$ meson $\raa$ at mid-rapidity provides a necessary baseline for further investigation of the longitudinal-dependent observables that rely on the titled initial condition of the bulk matter and the spacetime profiles of the external field.

\begin{figure}[tbp]
\begin{center}
\includegraphics[trim=0cm 0.2cm 0cm 0cm,width=8 cm,height=5 cm,clip]{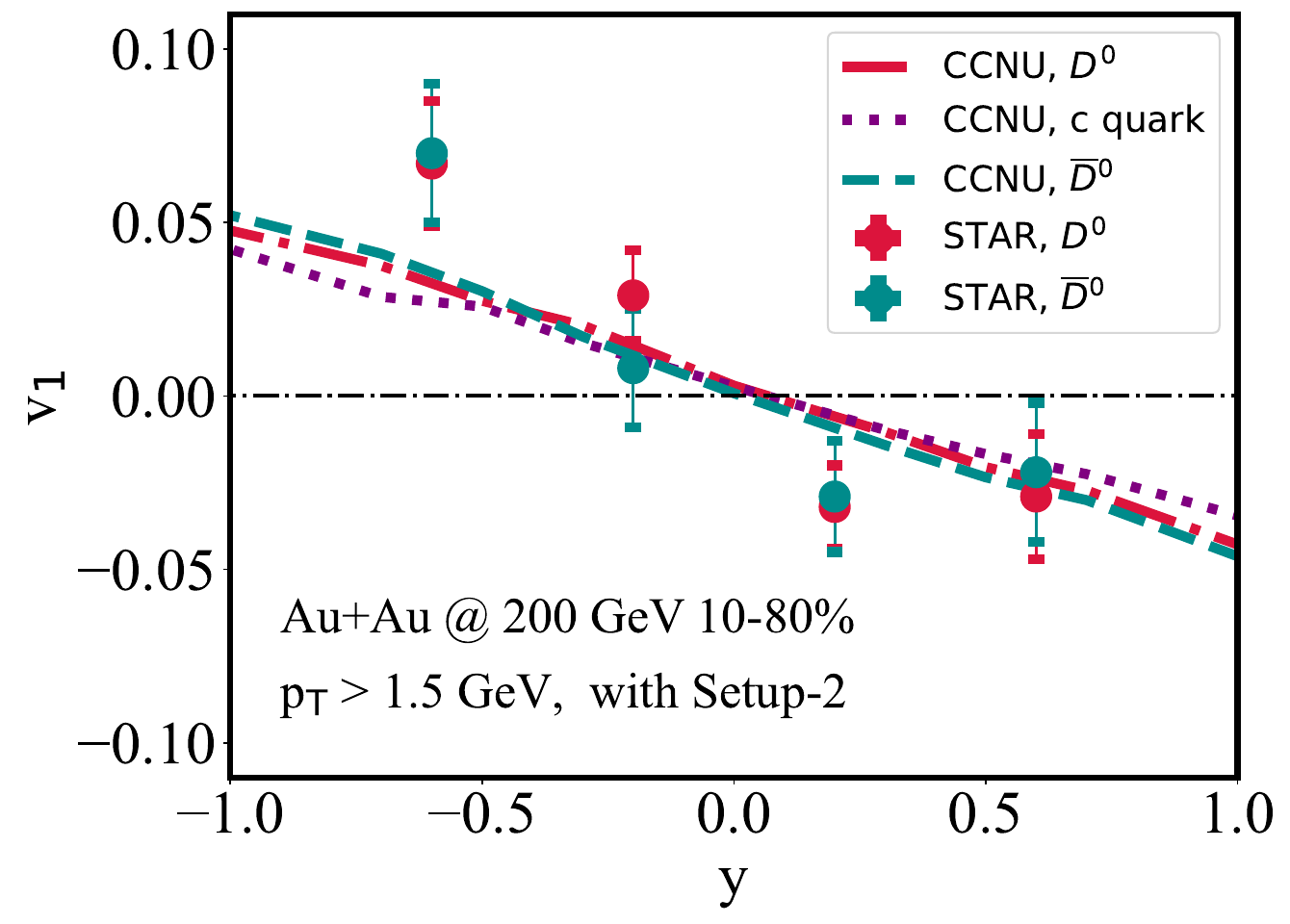}~\\
\includegraphics[trim=0cm 0.2cm 0cm 0cm, width=8 cm,height=5 cm,clip]{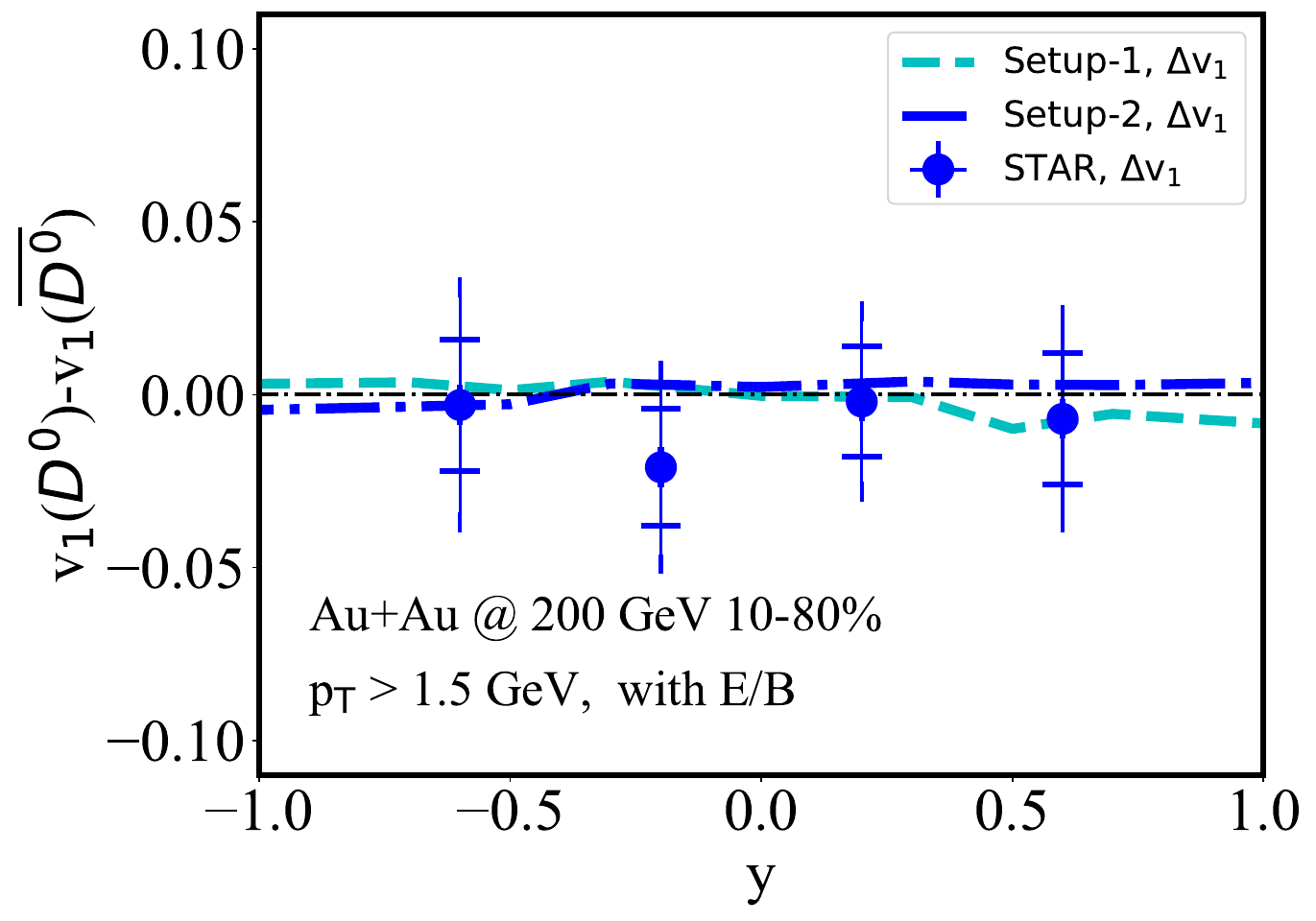}~
\end{center}
\caption{(Color online) Upper panel: directed flow of charm quarks, $D^{0}$ and $\dbar$ mesons as a function of rapidiy in 10\%-80\% Au+Au collisions at $\snn=200$~GeV with electromagnetic field setup-2. Lower panel: the direct flow splitting $\Delta v_{1} = v_{1}(D^{0})-v_{1}(\bar{D}^{0})$ compared between two setups of the electromagnetic field. Results are compared to the STAR data~\cite{STAR:2019clv}.}
\label{F1:RAA_V1_Au200}
\end{figure}

In Fig.~\ref{F1:RAA_V1_Au200}, we investigate the rapidity dependence of the $D$ meson $v_1$ in 10-80\% Au-Au collisions. Effects of both the titled initial condition and the electromagnetic field have been included. By using the second setup of the electromagnetic field described in Sec.~\ref{subsec:emfield}, we observe that both $D^0$ and $\dbar$ exhibit a negative slope of $v_1$ {\it vs.} $y$ in the upper panel of the figure. The directed flow $\dbar$ decreases slightly faster than that of $D^0$ because $\bar{c}$ and $c$ quarks are deflected towards different directions along the $x$-axis by the electromagnetic field. Note that according to Eq.~(\ref{eq:eE}), the negative $B_y$ with decaying magnitude results in positive $E_x$ at $z<0$ and negative $E_x$ at $z>0$. With such configuration, the electric and magnetic fields deflect a given charge into opposite directions. For instance, for a positive charge traveling along the $+z$ direction, the magnetic field deflects it towards $+x$ while the electric field deflects towards $-x$. In the end, whether $v_1$ of $D^0$ decreases faster or slower than that of $\dbar$ with respect to rapidity relies on the competing strength between electric and magnetic fields. In the upper panel, we also present the directed flow of charm quarks before hadronization. Comparing between results for $c$-quark and $D^0$, a weak effect from the hadronization process can be seen on the heavy flavor $v_1$.

For a closer investigation on the effect of electromagnetic field, we present the directed flow splitting $\dv=v_{1}(D^{0})-v_{1}(\dbar)$ in the lower panel of Fig.~\ref{F1:RAA_V1_Au200}, compared between the two field setups. However, due to the relatively weak magnitude of electromagnetic field at RHIC, compared to that at LHC as will be shown later, the $v_1$ splitting from both field setups are small. The evolution profile of the electromagnetic field is hard to be constrained using the RHIC data with the current large uncertainties. The unbalanced energy loss in $\pm x$ directions through a tilted medium (as shown in Fig.~\ref{fig:tiltedMedium}) is the main source of the heavy flavor $v_1$ at RHIC. This is consistent with the findings presented in Refs.~\cite{Chatterjee:2017ahy,Chatterjee:2018lsx,Beraudo:2021ont,Oliva:2020doe}.

\begin{figure}[tbp]
\begin{center}
\includegraphics[trim=0cm 0.2cm 0cm 0cm,width=8 cm,height=5 cm,clip]{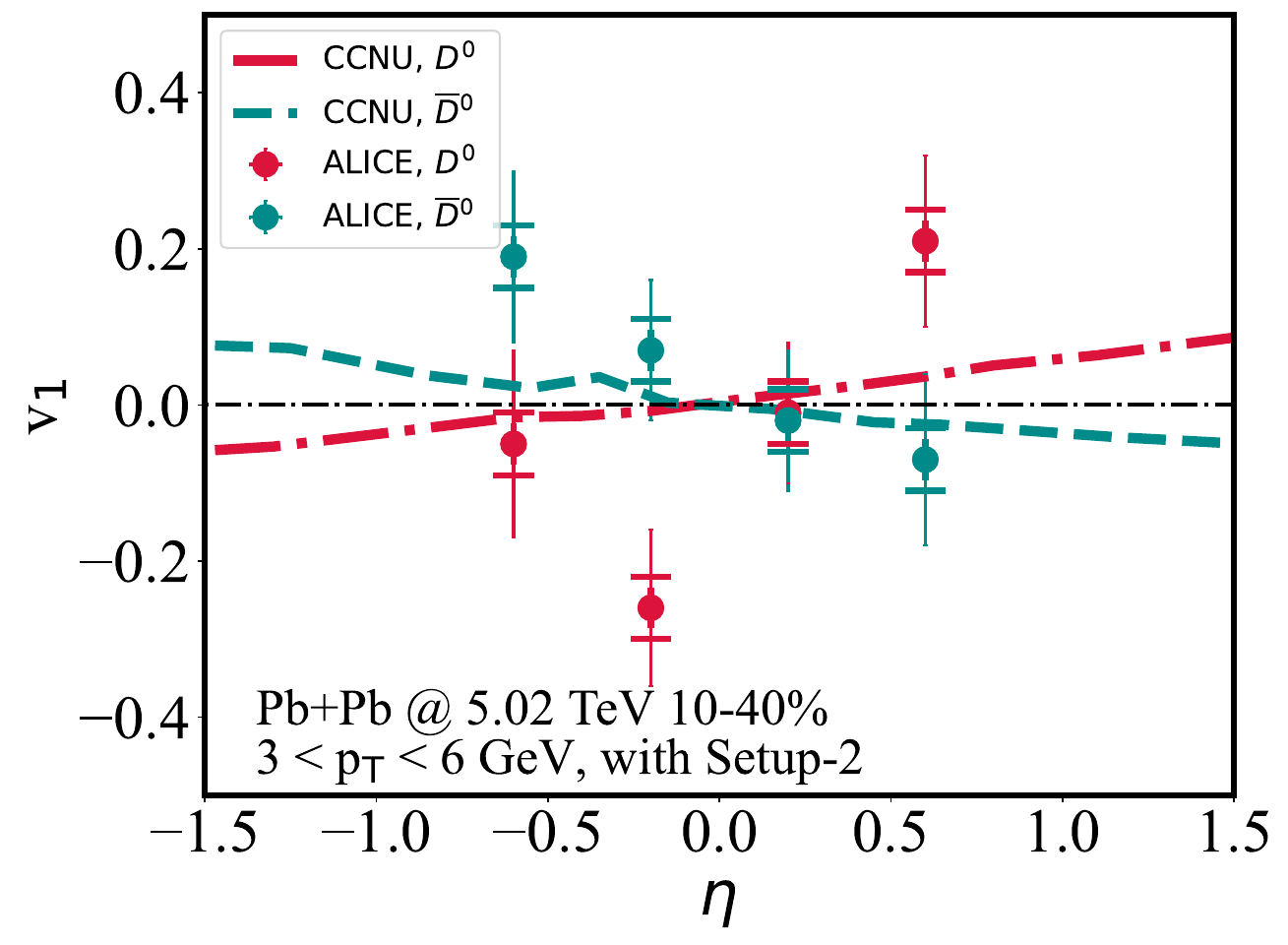}~\\
\includegraphics[trim=0cm 0.2cm 0cm 0cm,width=8 cm,height=5 cm,clip]{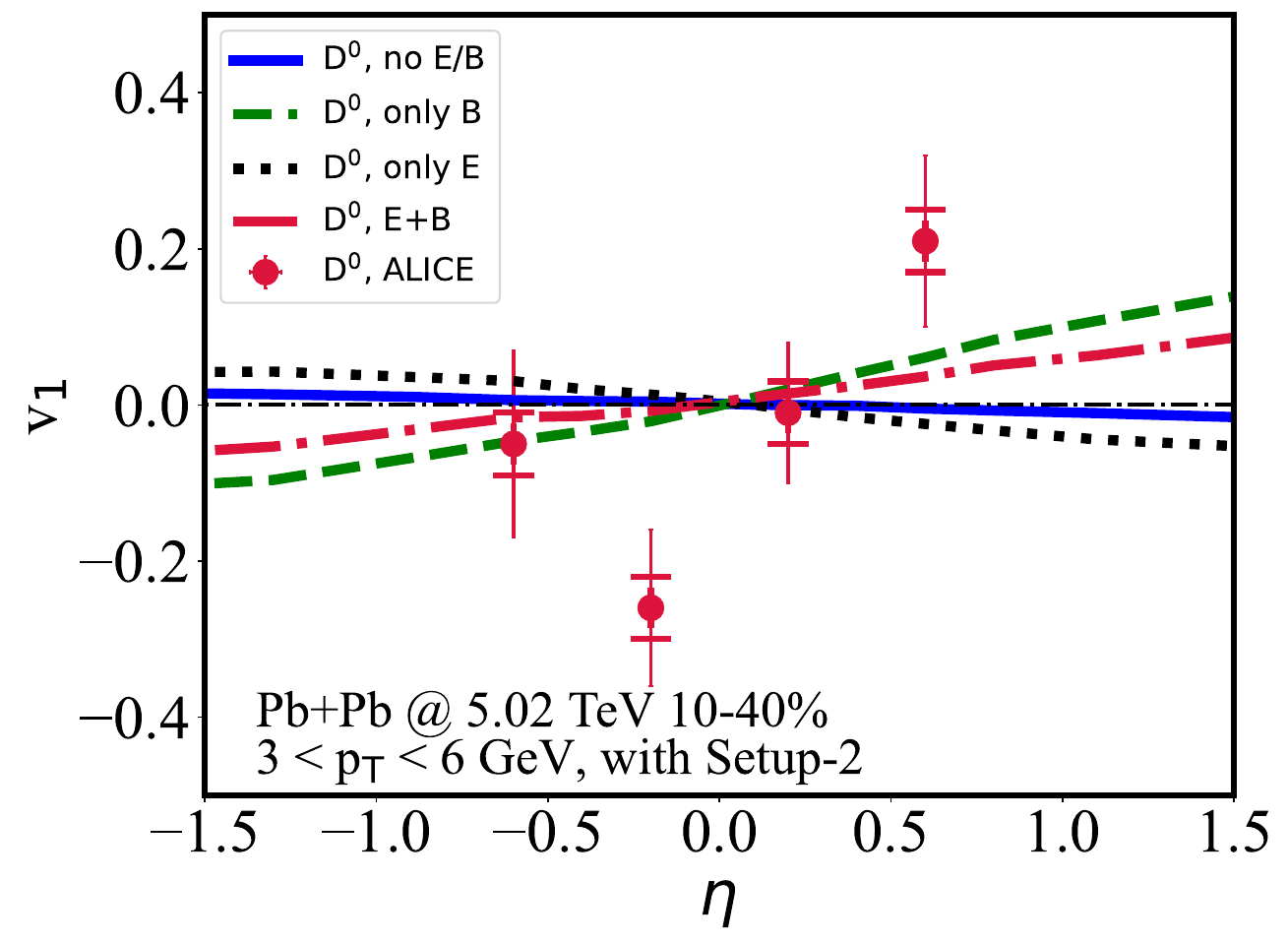}~\\
\includegraphics[trim=0cm 0.2cm 0cm 0cm,width=8 cm,height=5 cm,clip]{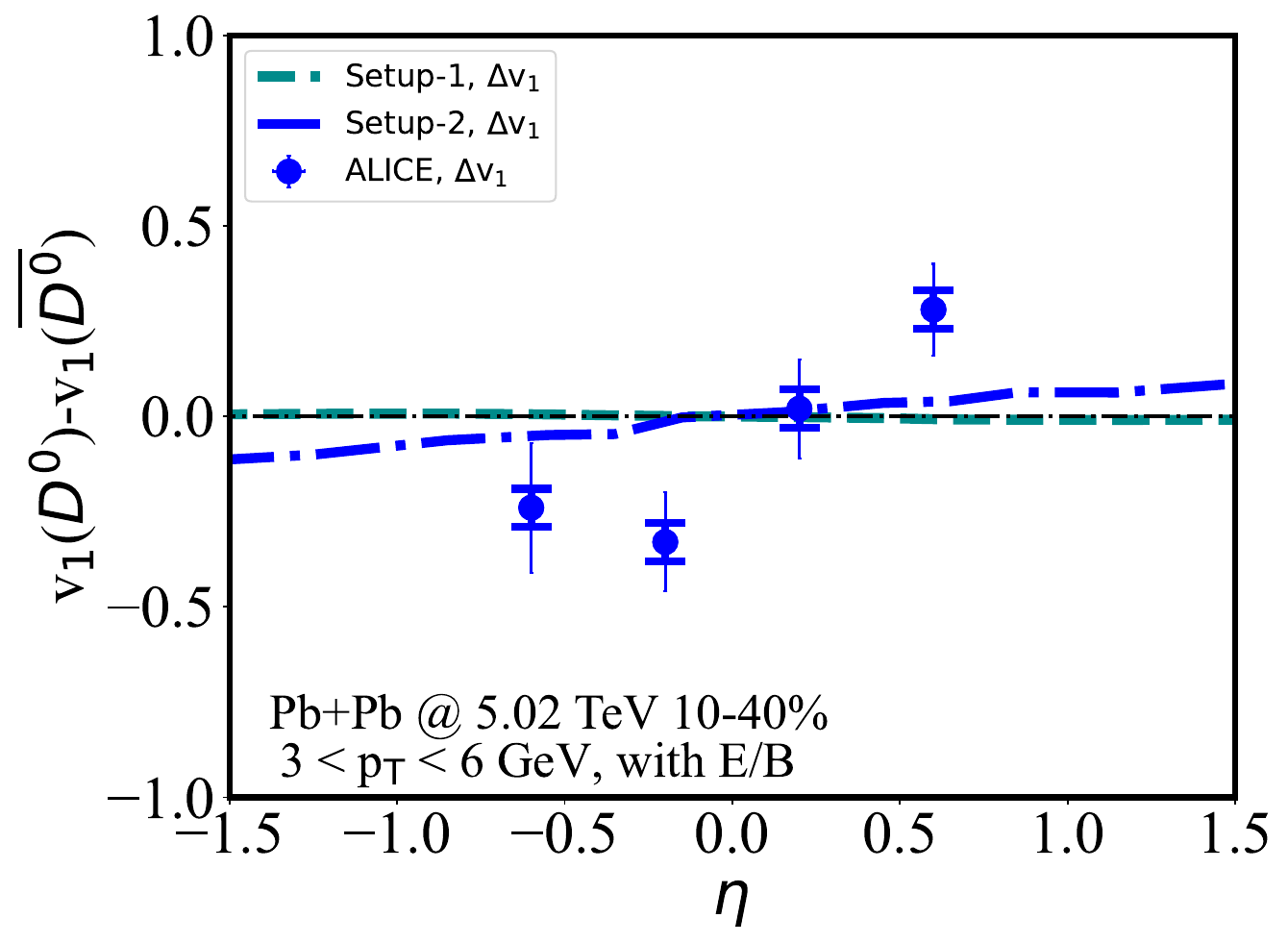}~
\end{center}
\caption{(Color online) Directed flow coefficients of $D^{0}$ and $\dbar$ mesons and their difference in 10\%-40\% Pb+Pb collisions at $\snn$ = 5.02 TeV, compared to the ALICE data~\cite{ALICE:2019sgg}.}
\label{F:PbPb5020D_v1}
\end{figure}

In Fig.~\ref{F:PbPb5020D_v1}, we further study the directed flow of $D$ mesons in 10\%-40\% Pb+Pb collisions. In the upper panel, with the setup-2 of electromagnetic field, one observes that while $\dbar$ shows a negative slope of $v_1$ with respect to pseudorapidity ($\eta$), $D^0$ shows the opposite. To understand this qualitative difference from the RHIC result, we separate different origins of $v_1$ in the middle panel. Without introducing the electromagnetic field (blue solid curve), the titled geometry of the QGP medium yields very small magnitude of $v_1$. This is due to the much more balanced initial condition between the forward and backward rapidity region in more energetic nuclear collisions at LHC than at RHIC. The weaker tilt of the bulk medium at LHC can also be reflected by the smaller $v_1$ of soft hadrons emitted from the QGP, as shown in our earlier study~\cite{Jiang:2021ajc}. On the other hand, the electromagnetic field is much stronger at LHC than at RHIC. Same as the earlier discussion, we observe the pure magnetic field (green dashed curve) leads to a positive slope of $v_1(\eta)$ for the positively charged charm quark and thus $D^0$. To the contrary, the pure electric field (black dotted curve) leads to a negative slope. Because of the larger magnitude of $B_y$ than $E_x$ within setup-2, the slope is still positive after electric and magnetic fields are combined (red dotted-dashed curve). This positive slope also overwhelms the small negative slope contributed by the tilted medium geometry (blue solid curve), resulting in a final positive slope for $D^0$ after all effects are combined.

In the lower panel of Fig.~\ref{F:PbPb5020D_v1}, the $\dv$ between $D^0$ and $\dbar$ is compared between our two setups. As illustrated in Fig.~\ref{f1:EBfield}, the magnitude of the electromagnetic field is much larger in setup-2 than in setup-1. Besides, setup-1 yields similar magnitudes between $B_y$ and $E_x$, but setup-2 provides a larger $B_y$ than $E_x$ during the entire QGP lifetime. As a result, one observes very small $\dv$ here from setup-1, while apparently larger $\dv$ (with a positive slope with respect to $\eta$) from setup-2. The ALICE data~\cite{ALICE:2019sgg} prefer the field profile modeled with setup-2. In the rest of this work, we will continue using setup-2 for predicting the observables of heavy flavor decay leptons.

\begin{figure}[tbp]
\begin{center}
\includegraphics[trim=0cm 0.2cm 0cm 0cm,width=8 cm,height=5 cm,clip]{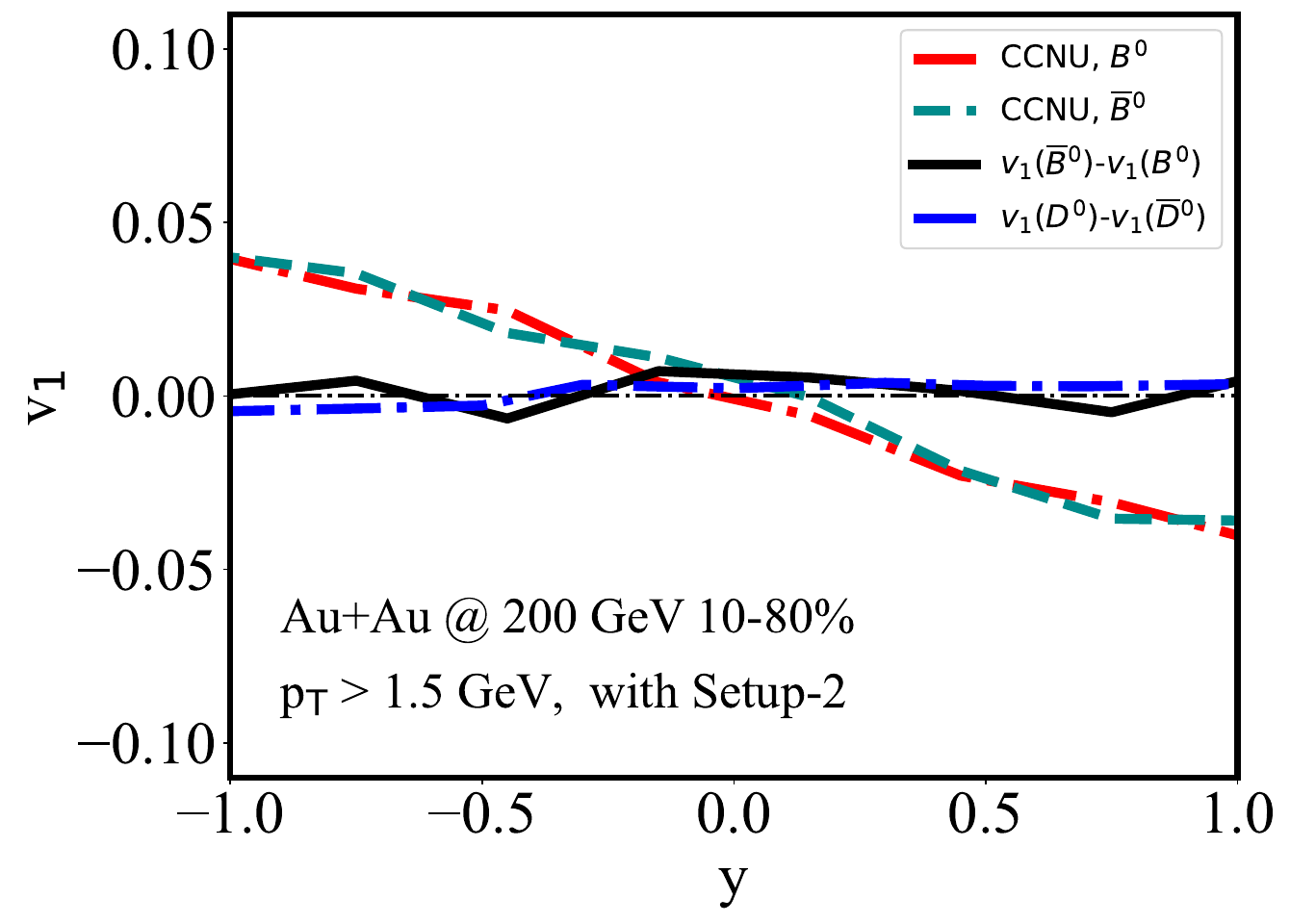}~\\
\includegraphics[trim=0cm 0.2cm 0cm 0cm,width=8 cm,height=5 cm,clip]{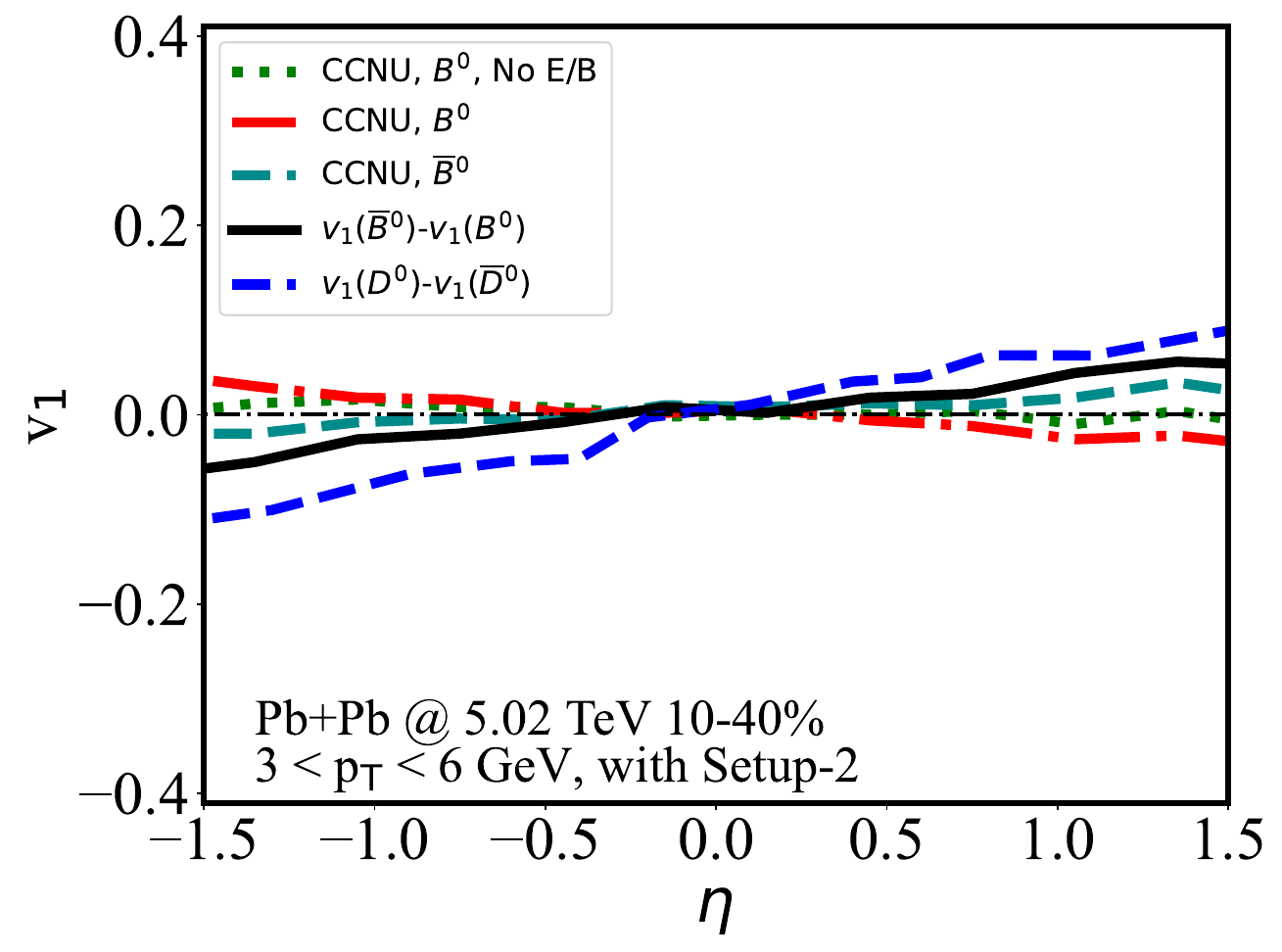}~
\end{center}
\caption{(Color online) Directed flow coefficients of $B^{0}$ and $\bar{B}^{0}$ mesons and their difference in 10-80\% 200 AGeV Au+Au collisions (upper panel) and 10-40\% 5.02 ATeV Pb+Pb collisions (lower panel).}
\label{fig:beauty}
\end{figure}

Similar investigations have also been implemented for $B$ mesons in Fig.~\ref{fig:beauty}, where we assume $b$-quarks share the same $D_\mathrm{s}$ with $c$-quarks, which is able to provide reasonable descriptions of the $B$ meson $R_\mathrm{AA}$ at RHIC and LHC. Consistent conclusions with $D$ mesons can be drawn here. In the upper panel, we observe a negative slope of $v_1$ as a function of rapidity at RHIC for both $\bar{B}^0$ and $B^0$, with no apparent difference between them. This indicates the negligible impact of electromagnetic field at RHIC, while the $B$ meson $v_1$ is mainly driven by the tilted geometry of the QGP medium. The $v_1$ difference $v_1(\bar{B}^0)-v_1({B}^0)$ is also compared with  $v_1({D}^0)-v_1(\bar{D}^0)$: they are equally small. In the lower panel, we observe different slopes between $\bar{B}^0$ (positive) and ${B}^0$ (negative) at LHC, indicating the dominant effect from the electromagnetic field from the more energetic collisions at LHC. However, the $v_1$ splitting between $\bar{B}^0$ and ${B}^0$ is much smaller than that between ${D}^0$ and $\bar{D}^0$, which results from the smaller electric charge carried by $b$-quarks ($-1/3$) compared to $c$-quarks ($2/3$), and the larger mass of $b$-quarks (or smaller velocity for a given momentum) than $c$-quarks, both reducing the Lorentz force experienced by $b$-quarks than $c$-quarks.

\subsection{$\raa$ and $v_{1}$ of charm decay electrons}

Measurements of open heavy flavor at large rapidity have been carried out via detecting electrons and muons from the decay of charm and beauty hadrons using the electron/muon spectrometer in nuclear collision experiments.
The new Muon Forward Tracker in the LHC Run 3 is able to separate muons from charm and beauty semi-leptonic decays~\cite{ATLAS:2020yxw}, allowing a clean investigation on the properties heavy quarks with particular species.
These upgraded experiments provide a broad coverage of rapidity, making heavy flavor decay electrons and muons ideal candidates for probing the tilted medium geometry with respect to the longitudinal direction, as well as the the evolution profile of the electromagnetic field.

\begin{figure}[tbp]
\begin{center}
\includegraphics[trim=0cm 0.2cm 0cm 0cm,width=8 cm,height=5 cm,clip]{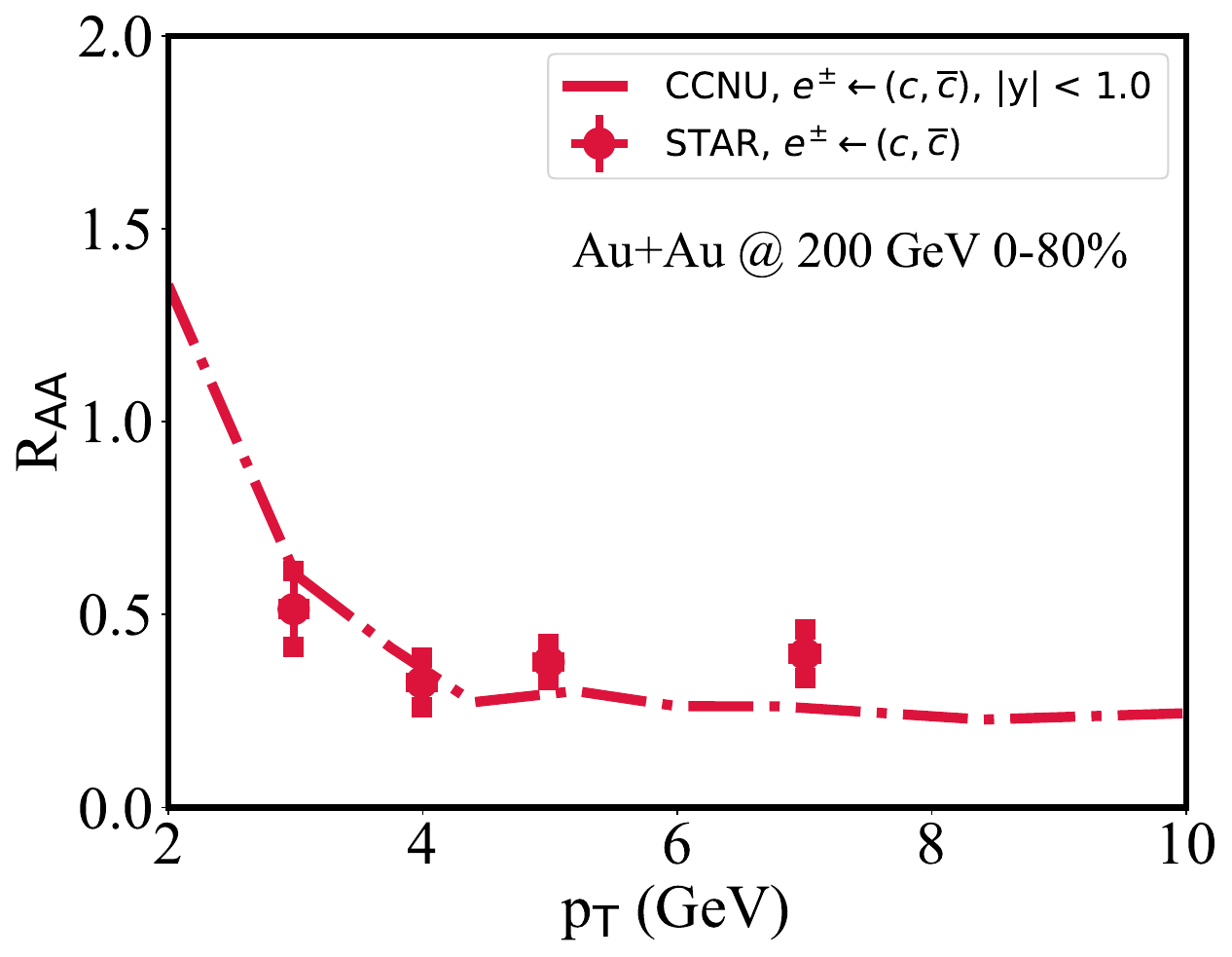}~\\
\includegraphics[trim=0cm 0.2cm 0cm 0cm,width=8 cm,height=5 cm,clip]{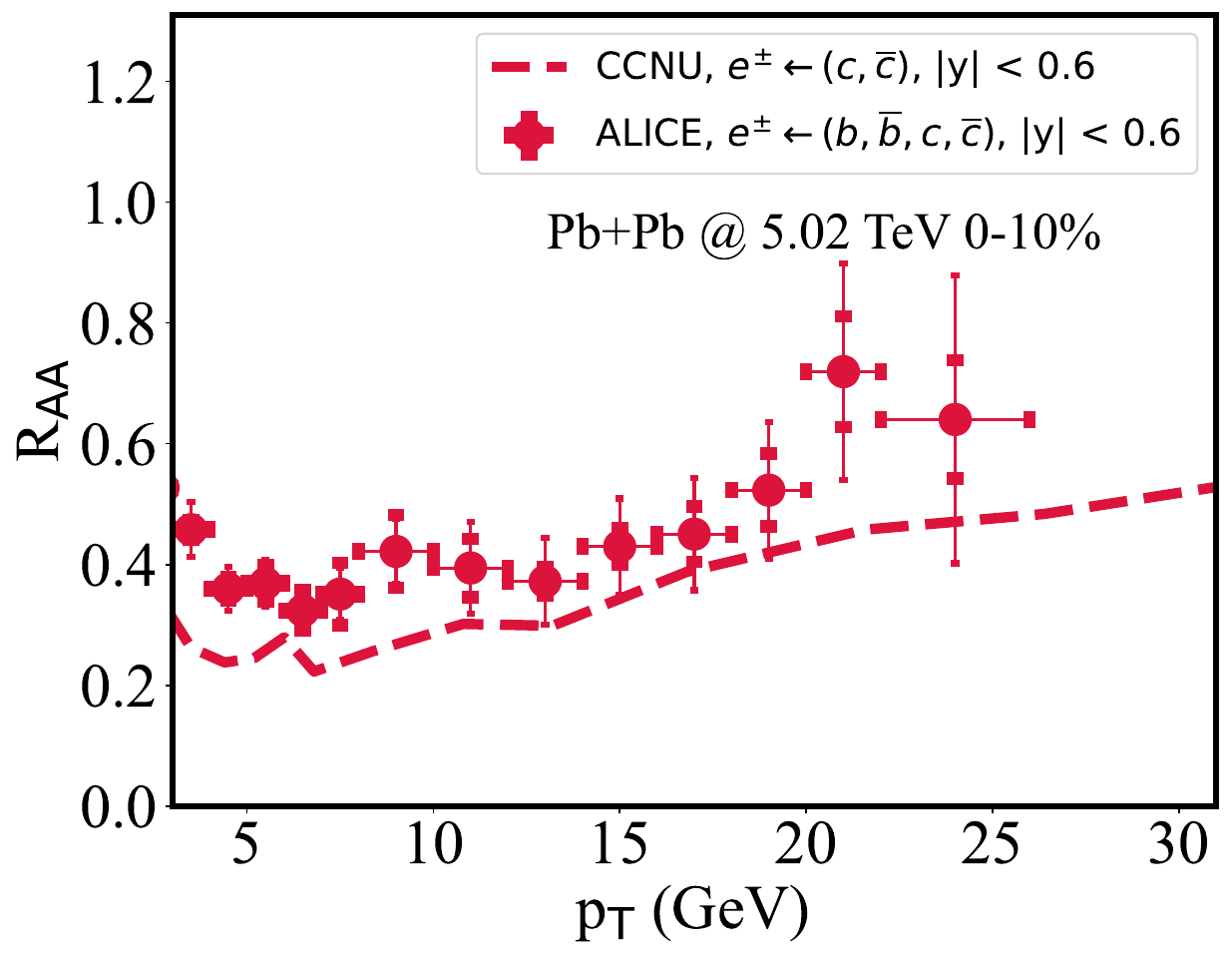}~
\end{center}
\caption{(Color online) Nuclear modification factor of charm decay electrons in 0-80\% Au+Au collisions at $\snn=200$~GeV (upper panel) and 0-10\% Pb+Pb collisions at $\snn=5.02$~TeV (lower panel), compared to the STAR data~\cite{Licenik:2020cjc,Kelsey:2020bms} and the ALICE data~\cite{ALICE:2019nuy} respectively.}
\label{F:raa_electron}
\end{figure}

We start with validating our model calculation with the $R_\mathrm{AA}$ of charm decay electrons at RHIC and LHC. As shown in Fig.~\ref{F:raa_electron}, with the same transport calculations for $D$ mesons in the previous subsection, our $\raa$ of charm decay electrons agrees with the STAR data for 0-80\% Au+Au at $\snn=200$~GeV (upper panel) at mid-rapidity. For 0-10\% Pb+Pb collisions at $\snn=5.02$~TeV (lower panel), since the current ALICE measurement includes contributions from both charm and beauty decay electrons, our result with contribution from charm quarks alone is expected to be a little smaller than the data, considering the stronger energy loss experienced by lighter charm quarks than heavier beauty quarks.

\begin{figure}[tbp]
\begin{center}
\includegraphics[trim=0cm 0.2cm 0cm 0cm,width=8 cm,height=5 cm,clip]{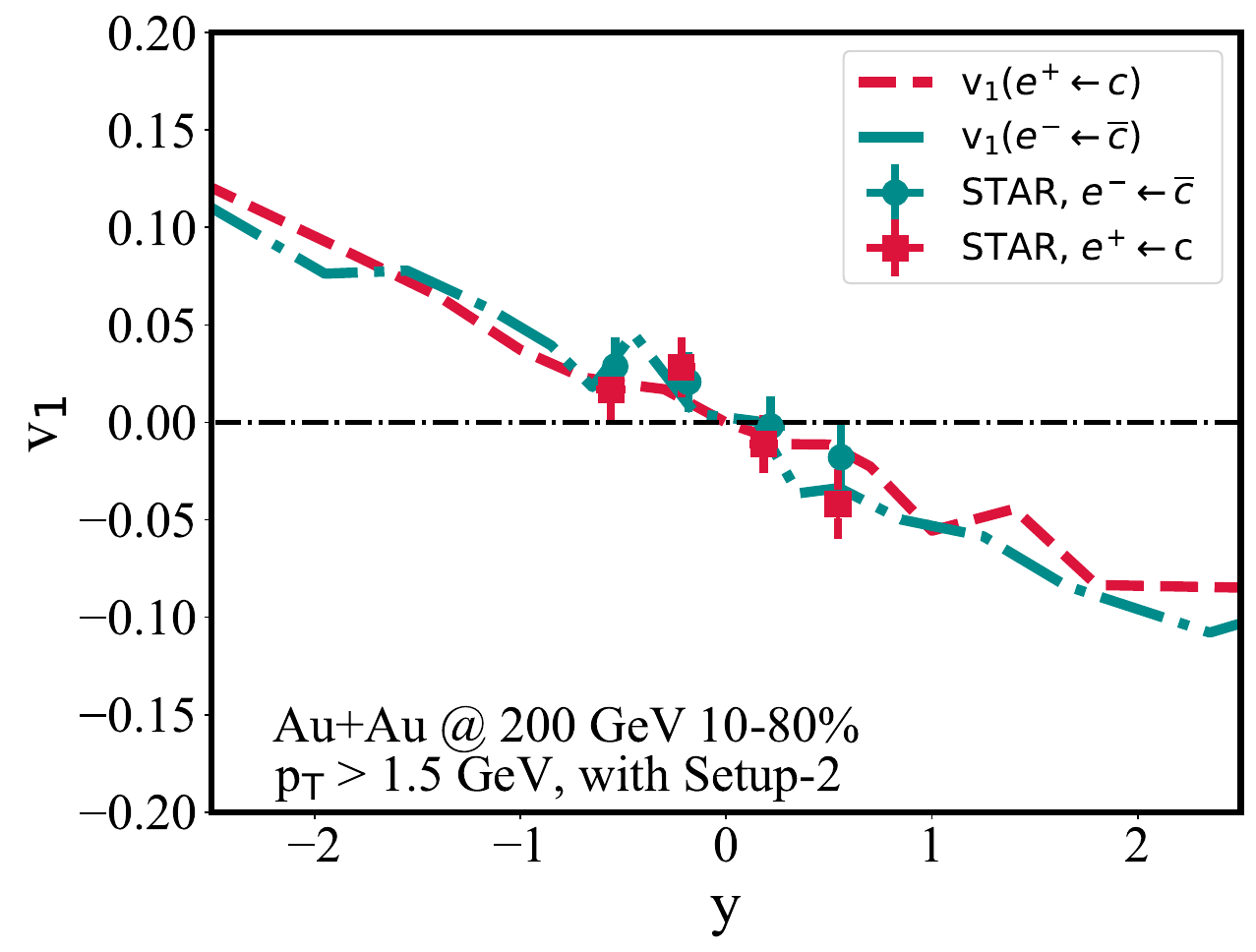}~\\
\includegraphics[trim=0cm 0.2cm 0cm 0cm,width=8 cm,height=5 cm,clip]{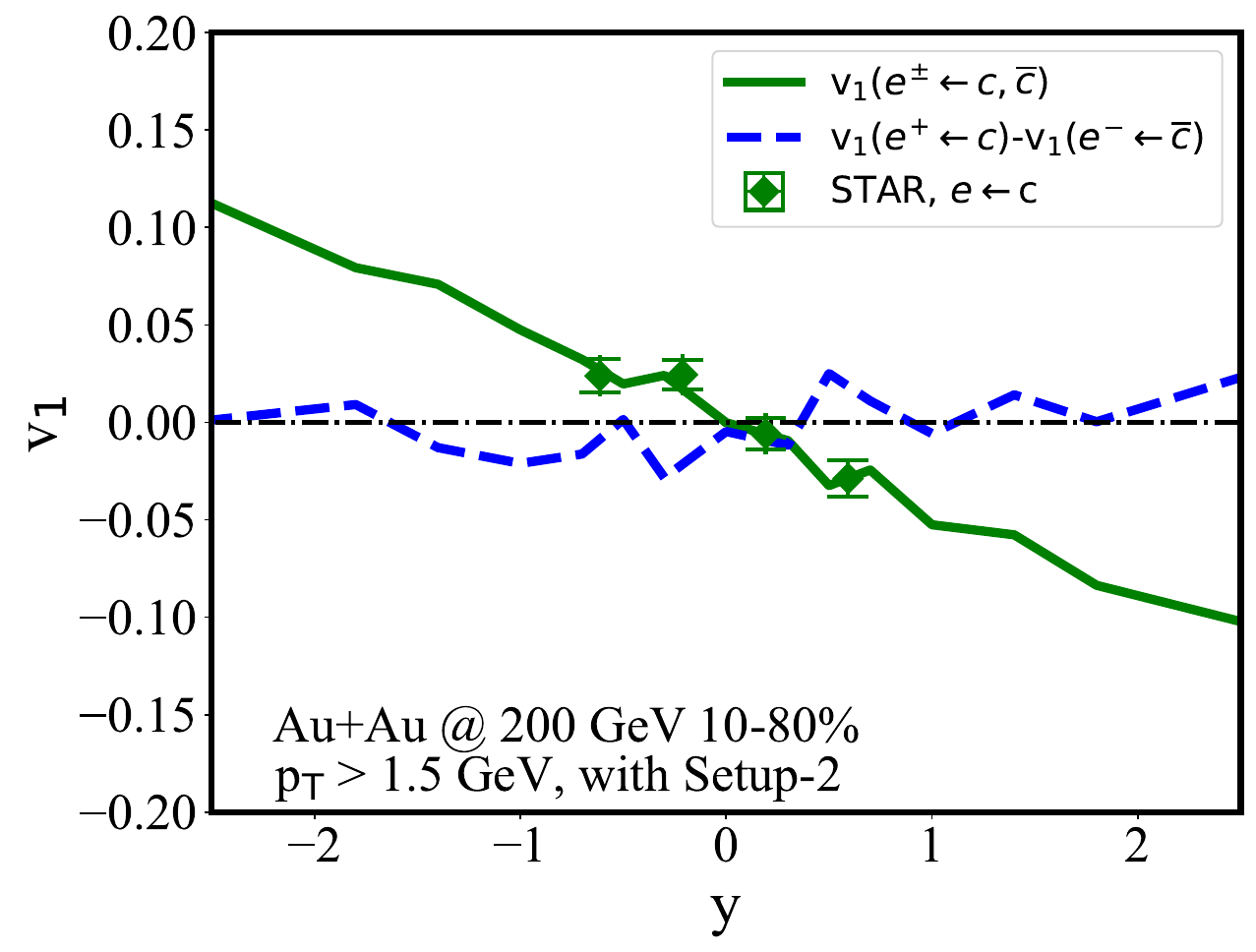}~\\
\includegraphics[trim=0cm 0.2cm 0cm 0cm,width=8 cm,height=5 cm,clip]{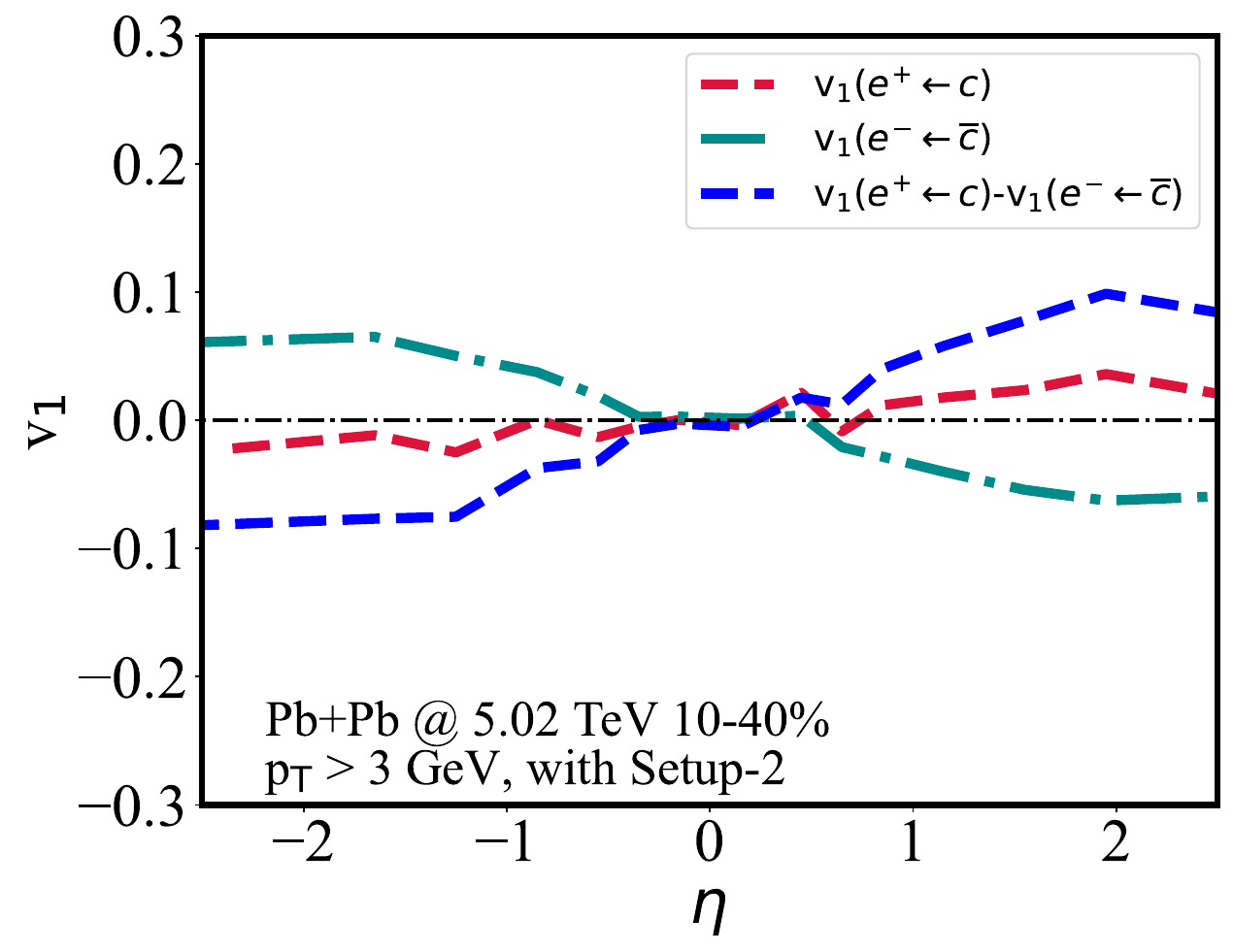}~
\end{center}
\caption{(Color online) Directed flow coefficient of charm decay electrons in 10-80\% Au+Au collisions at $\snn=200$~GeV, compared to the STAR data~\cite{Licenik:2020cjc,Kelsey:2020bms,Kramarik:2021emg}, and in 10-40\% Pb+Pb collisions at $\snn=5.02$~TeV.}
\label{F:v1_electron}
\end{figure}

The directed flow coefficients of charm decay electrons are then presented in Fig.~\ref{F:v1_electron}, with setup-2 adopted for the electromagnetic field. As shown in the upper panel, the rapidity dependence of the charm decay electron $v_1$ agrees with the STAR data in 10-80\% Au+Au collisions at $\snn=$ 200 GeV, with a slope parameter extracted as $dv_{1} /dy = - 0.045\pm0.005$ around the $y=\pm 1$ regions. Little difference can be observed between $c$-decay $e^+$ and $\bar{c}$-decay $e^-$. In the middle panel, we present the average $v_1$ of $c$-decay $e^+$ and $\bar{c}$-decay $e^-$, compared to the difference between them on the same scale. One observes that the difference, resulting from the electromagnetic effect, is much smaller than the average, resulting from the tilted geometry of the QGP medium. This confirms the asymmetric medium profile is the dominant origin of the heavy flavor $v_1$ at RHIC, consistent with the findings using the $D$ meson $v_1$ in the previous subsection.

Shown in the lower panel of Fig.~\ref{F:v1_electron} is our prediction for $v_{1}$ of charm decay electrons in 10\%-40\% Pb+Pb collisions at $\snn=5.02$~TeV. Separate results for $c$-decay positrons and $\bar{c}$-decay electrons, together with their difference, are presented. While $v_{1}(e^{-}\leftarrow \bar{c})(\eta)$ shows a negative slope, $v_{1}(e^{+}\leftarrow c)(\eta)$ shows a positive slope, indicating the stronger electromagnetic effect on the heavy flavor $v_1$ than the geometric effect of the medium at LHC. The $v_1$ splitting between positron and electron increases with $\eta$, whose slope parameter is extracted as $d\dv /d\eta$ = 0.05$\pm$0.01 around $y=\pm 1$, which can be tested by future measurement at LHC.

\subsection{$\raa$ and $v_{1}$ of charm decay muons}

Finally, we close our study with the prediction for heavy-flavor decay muons, which is the only probe so far of heavy flavor dynamics at forward rapidity in nuclear collisions. Contributions from charm and beauty quarks are also able to be distinguished at ATLAS within the $|y|<2.0$ range~\cite{ATLAS:2020yxw}.

\begin{figure}[tbp]
\begin{center}
\includegraphics[trim=0cm 0.2cm 0cm 0cm,width=8 cm,height=5 cm,clip]{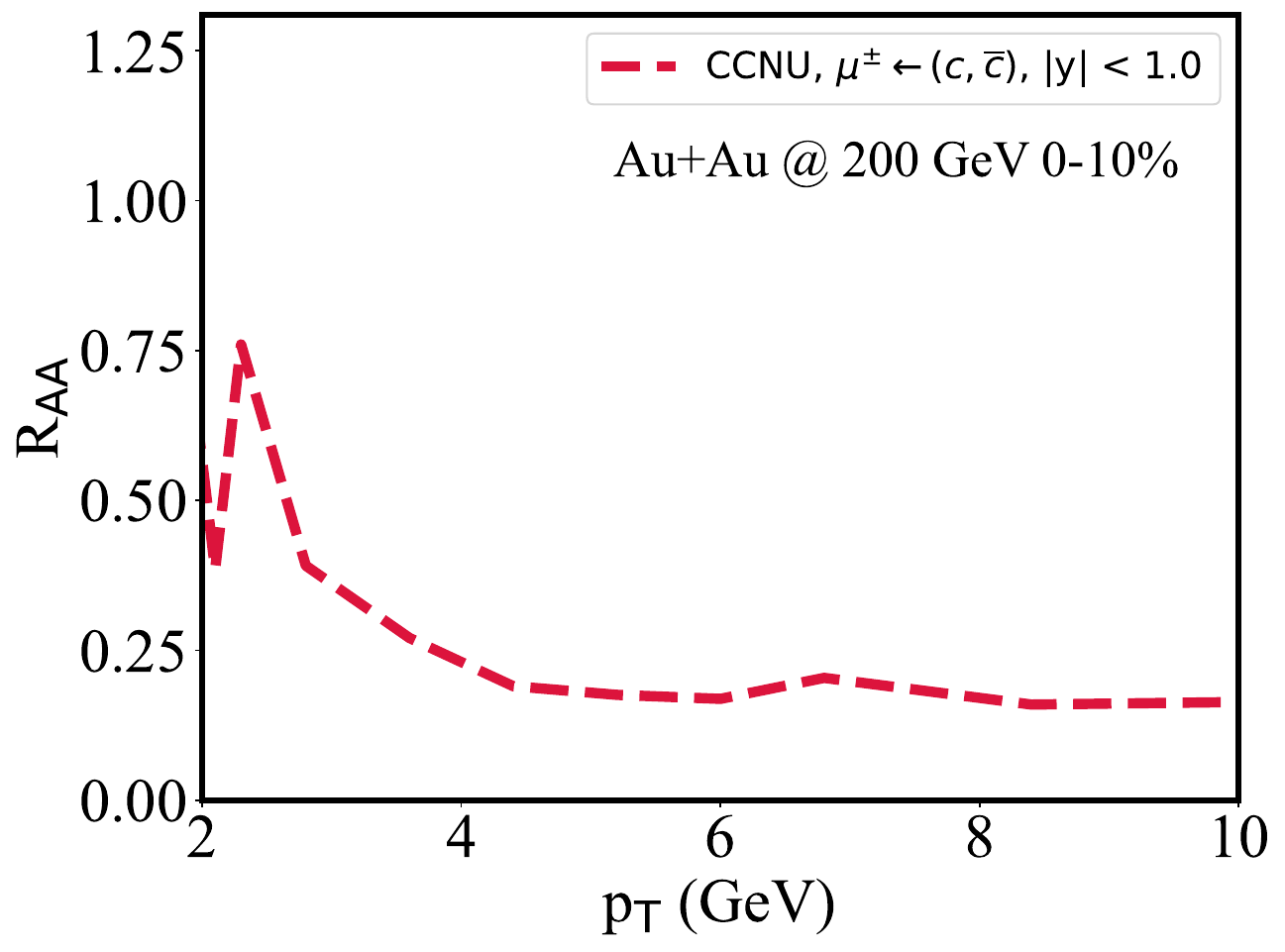}~\\
\includegraphics[trim=0cm 0.2cm 0cm 0cm,width=8 cm,height=5 cm,clip]{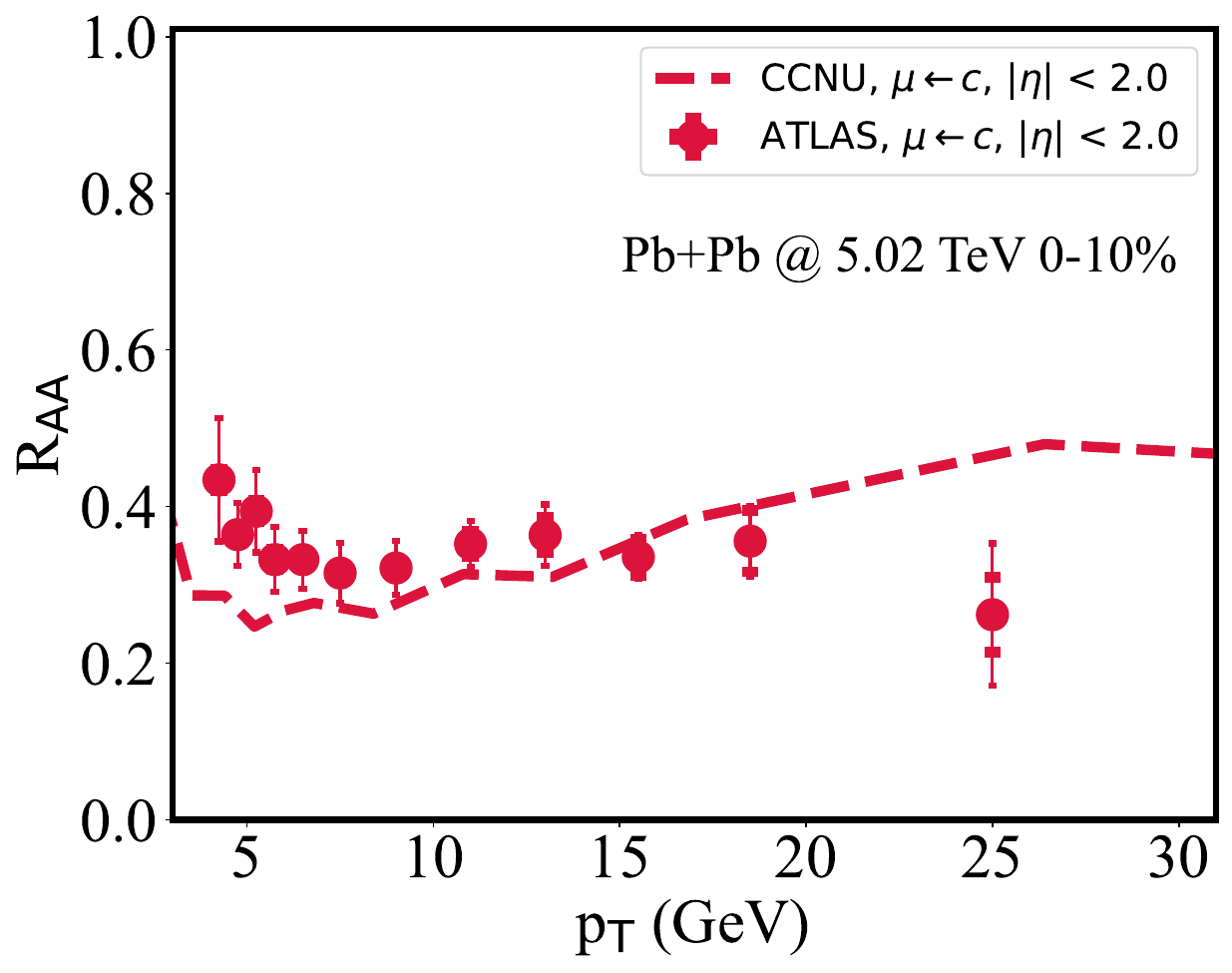}~
\end{center}
\caption{(Color online) Nuclear modification factor of muons in 0-10\% Au+Au collisions at $\snn=200$~GeV and Pb+Pb collisions at $\snn=5.02$~TeV. The latter is compared to the ATLAS data~\cite{ATLAS:2020yxw,ATLAS:2021xtw}.}
\label{F:raa_muons}
\end{figure}

Displayed in Fig.~\ref{F:raa_muons} is the $\raa$ of charm decay muons in 0-10\% Au+Au collisions at $\snn = 200$~GeV (upper panel) and Pb+Pb collisions at $\snn= 5.02$~TeV (lower panel). Our calculation for the latter is in reasonable agreement with the available data from the ATLAS collaboration in the mid-rapidity region~\cite{ATLAS:2020yxw}. The $\raa$ of charm decay muons should be close to that of charm decay electrons previously shown in Fig.~\ref{F:raa_electron}. The residual difference should come from the decay functions that takes different fractions of momentum from the parent charm quarks for electrons and muons.

\begin{figure}[tbp]
\begin{center}
\includegraphics[trim=0cm 0.2cm 0cm 0cm,width=8 cm,height=5 cm,clip]{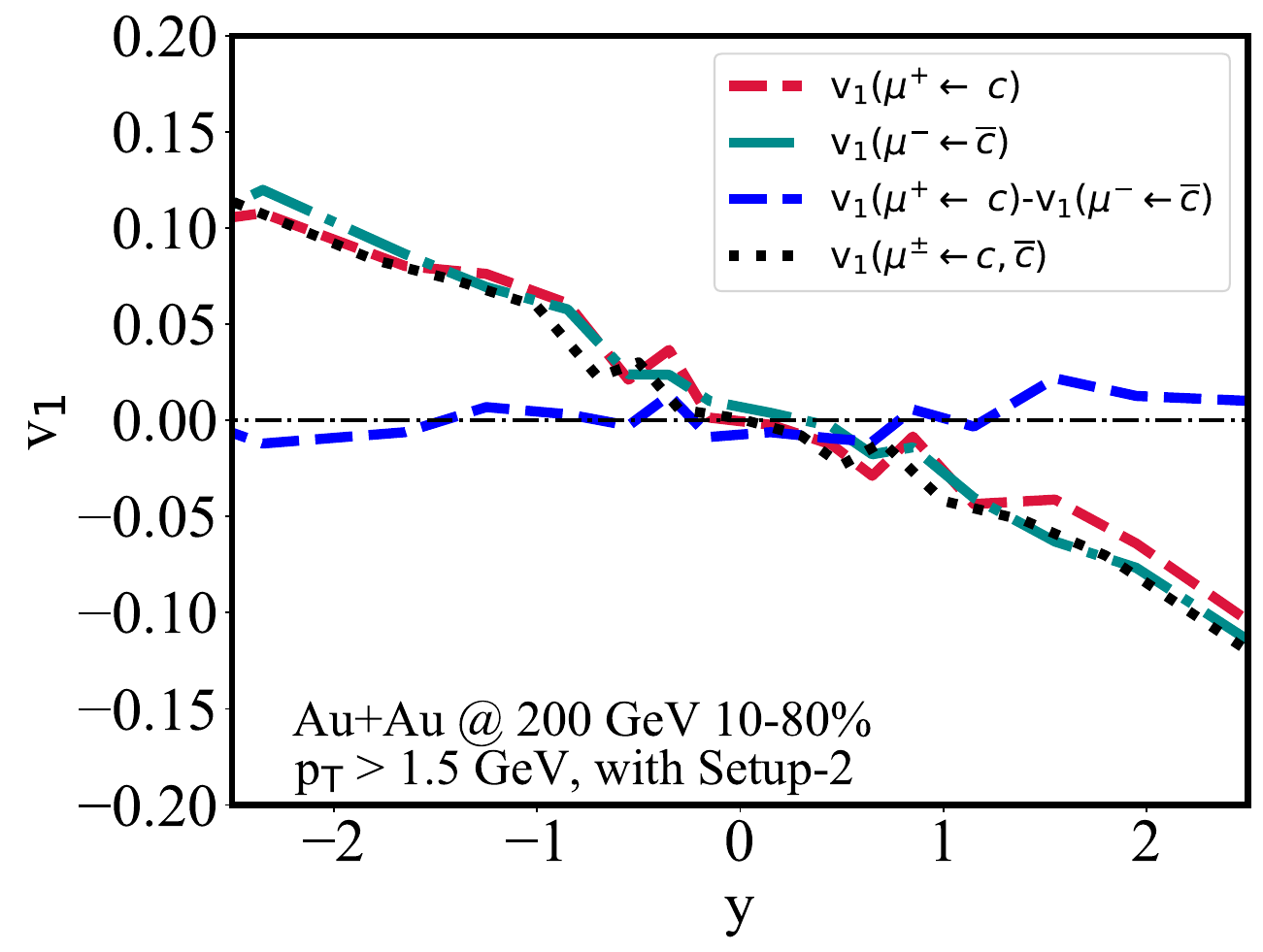}~\\
\includegraphics[trim=0cm 0.2cm 0cm 0cm,width=8 cm,height=5 cm,clip]{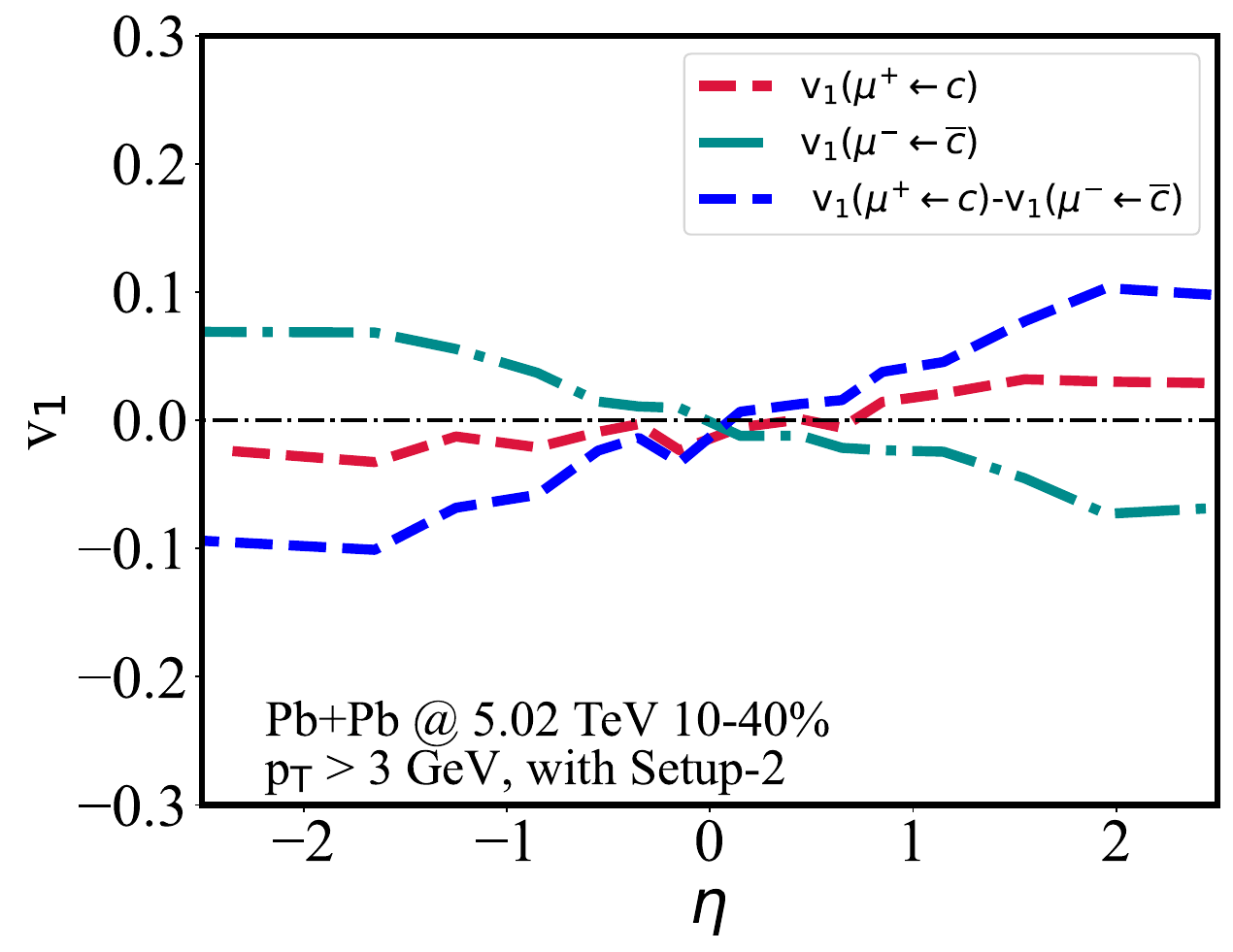}~
\end{center}
\caption{(Color online) Directed flow coefficient of charm decay muons in 10-80\% Au+Au collisions at $\snn=200$~GeV and 10-40\% Pb+Pb collisions at $\snn=5.02$~TeV.}
\label{F:v1_muons}
\end{figure}

The directed flow coefficient of charm decay muons are presented in Fig.~\ref{F:v1_muons}, upper panel for 10-80\% Au+Au collisions at $\snn = 200$~GeV and lower panel for 10-40\% Pb+Pb collisions at $\snn = 5.02$~TeV. Consistent with the previous results for charm decay electrons, in the upper panel, we observe the directed flow coefficients of $c$-decay $\mu^+$, $\bar{c}$-decay $\mu^-$, and their average almost overlap each other. The slope parameter for $v_1(\eta)$ is $dv_{1} /dy = -0.05\pm 0.005$ around $y=\pm 1$. Meanwhile, $\dv$ between $\mu^+$ and $\mu^-$ is close to zero. This further confirms the dominant contribution to $v_1$ from the tilted medium geometry at RHIC. Contrarily, the strong electromagnetic field dominates the formation of muon $v_1$ at LHC. As shown in the lower panel, while the slope of $v_1(\eta)$ is negative for $\bar{c}$-decay $\mu^-$, it is flipped for $c$-decay $\mu^+$ due to opposite Lorentz force on opposite charges. Their difference -- $v_1(\mu^+\leftarrow c)-v_1(\mu^-\leftarrow \bar{c})$ -- then increases with $\eta$, with a slope parameter $d\dv/d\eta = 0.07\pm 0.005$ extracted around $\eta=\pm 1$.

\section{Summary and outlook}
\label{section5}

We have conducted a systematic investigation on the interplaying mechanisms behind the heavy flavor directed flow $v_{1}$ and $\dv$. Effects from the titled initial condition of the bulk medium, the electric field and the magnetic field have been analyzed in detail within a modified Langevin transport model coupled to a (3+1)-D viscous hydrodynamic model CLVisc, which has been validated with the nuclear modification factor ($\raa$) data of both heavy mesons and their decay leptons.

We have illustrated, for the first time, that the tilted initial energy density profile is the main source of the heavy flavor $v_{1}$ at the RHIC energy, while the electromagnetic field dominates the $v_{1}$ formation at the LHC energy. Comparing between our two setups of the electromagnetic field, we have found that the splitting of $v_1$ between positive and negative charges ($\dv$) is sensitive to the decay speed and the relative strength between electric and magnetic fields. While the electric field results in a negative slope of the $v_1(y)$ function of positive charges, the magnetic field yields a positive slope. Compared to our setup-1 (the direct solution of the Maxwell equation), our setup-2 adopted from Ref.~\cite{Sun:2020wkg} provides a much slower decay of the field, while a stronger strength of the magnetic field than the electric field throughout the QGP lifetime, which is favored by the experimental data on the pseudorapidity dependence of $v_1$ at LHC. These have been consistently confirmed with our calculations across $D$ mesons and their decay electrons and muons. Our numerical results agree with the available data at RHIC and LHC, and await test with future electron measurement at LHC and muon measurement at RHIC and LHC. These new observations at RHIC are expected to place a more stringent constraint on the initial energy density profile of the QGP; while observations at LHC are expected to help refine our knowledge on the spacetime evolution of the electromagnetic field in energetic nuclear collisions.


Our current study on the directed flow $v_1$ and the directed flow splitting $\dv$ of heavy flavor can be further extended in several directions. For instance, a simultaneous investigation on the $\raa$ and $v_1$ ($\dv$) of heavy quarks in isobaric $^{96}_{44}$Ru+$^{96}_{44}$Ru collisions and $^{96}_{40}$Zr+$^{96}_{40}$Zr collisions at RHIC and O+O and Ar+Ar collisions at LHC could provide additional constraints on the system size dependence of the tilted initial condition of the QGP, as well as the evolution profiles of the electromagnetic fields.
While a qualitative agreement can be achieved between our model calculation with setup-2 of electromagnetic field and the experimental data, a more precise quantitative agreement would requires a much more dedicate calculation of the electromagnetic field. Recent studies~\cite{Li:2016tel,Siddique:2022ozg} have shown that including electric and chiral magnetic conductivities can affect both the decay speed and the spatial symmetry of the electromagnetic field, which may also influence the azimuthal distribution of heavy quarks and their decay products. These will be incorporated in our model calculation in an upcoming effort.

\begin{acknowledgements}
We are grateful for helpful discussions with Chun Shen, Jiaxing Zhao, Yu-Fei Liu, Yifeng Sun, 
Xiaowen Li and Guang-You Qin. This work was supported by the National Natural Science Foundation of China (NSFC) under Grant Nos.~11935007, 12175122 and 2021-867, Guangdong Major Project of Basic and Applied Basic Research No.~2020B0301030008, the Natural Science Foundation of Hubei Province No.~2020CFB864,~2021CFB272, the Education Department of Hubei Province of China with Young Talents Project No.~Q20212703, the open foundation of Key Laboratory of Quark and Lepton Physics (MOE) No.~QLPL2021P01 and
the Xiaogan Natural Science Foundation under Grant No.~XGKJ2021010016. Computational resources were provided by the Center of Scientific Computing
at the Department of Physics and Electronic-Information Engineering, Hubei Engineering University.
\end{acknowledgements}

\bibliographystyle{unsrt}
\bibliography{heavy_v1ref}

\end{document}